\documentclass[12pt,a4paper]{article}

\usepackage[english]{babel}
    \addto\captionsenglish{}
\usepackage[autostyle=true]{csquotes}
\usepackage{amsmath,amssymb,amsfonts,pifont,mathtools}
\usepackage{bm}
\usepackage{xcolor}
\usepackage{amsthm}
\usepackage{graphicx}
\usepackage[caption=false,font=footnotesize]{subfig}
\usepackage{textcomp}
\usepackage{multirow}
\usepackage{tabularx}
\usepackage{booktabs}
\usepackage{threeparttable}
\usepackage{flushend}
\usepackage{microtype}
\usepackage[version=4]{mhchem}

\usepackage{geometry}
\usepackage[backend=biber,defernumbers=true,sorting=nty,maxbibnames=99]{biblatex}

\usepackage{siunitx}[=v2]
    \DeclareSIUnit{\unit}{U}
    \DeclareSIUnit{\px}{pixel}
\usepackage{titlesec}
\titleclass{\subsubtag}{straight}[\subsection]
\newcounter{subsubtag}

\titleformat{\subsubtag}[runin]{\normalfont\normalsize\bfseries}{\thesubsubtag}{1em}{}[:]

\titlespacing*{\subsubtag}{0pt}{2.5ex plus 0.5ex minus .1ex}{1ex plus .1ex}
\usepackage[algoruled,vlined]{algorithm2e}

\title{WAVEFRONT SUPER-RESOLUTION FOR ADAPTIVE OPTICS SYSTEM ON GROUND-BASED TELESCOPES}
\author{Yutong Wu, Roland Wagner, Raymond Chan, Ronny Ramlau}

\begin{document}
\maketitle

\begin{abstract}
In ground-based astronomy, Adaptive Optics (AO) is a pivotal technique, engineered to correct wavefront phase distortions and thereby enhance the quality of the observed images. Integral to an AO system is the wavefront sensor (WFS), which is crucial for detecting wavefront aberrations from guide stars, essential for phase calculations. Many models based on a single-WFS model have been proposed to obtain the high-resolution phase of the incoming wavefront. In this paper, we delve into the realm of multiple WFSs within the framework of state-of-the-art telescope setups for high-resolution phase reconstruction. We propose a model for reconstructing a high-resolution wavefront from a sequence of wavefront gradient data from multiple WFSs in a multi-frame post-processing setting. Our model is based on the turbulence statistics and the Taylor frozen flow hypothesis, incorporating knowledge of the wind velocities in atmospheric turbulence layers. We also introduce an $H_2$ regularization term, especially for atmospheric characteristics under von Karman statistics, and provide a theoretical analysis for $H^2$ space within $H^{11/6}$. Numerical simulations are conducted to demonstrate the robustness and effectiveness of our regularization term and multi-WFS reconstruction strategy under identical experimental conditions.
\end{abstract}

\section{Introduction}
In ground-based astronomy, images of objects in outer space are acquired by ground-based telescopes. However, the Earth's atmosphere consists of regions with varying temperatures and densities, causing a different index of refraction \cite{wyngaard1992atmospheric}. As light passes through these regions, it gets distorted, resulting in blurred images on the telescope. This image formation can be modeled mathematically as
\begin{equation}
    \label{eq: gkf}
    g(x,y) = \int_{\mathbb{R}^2} k(x-\xi, y-\eta)f(\xi,\eta) \,d(\xi,\eta),
\end{equation}
where $f(x,y)$ is the actual object being viewed in outer space, $g(x,y)$ is the observation of $f(x,y)$ captured by a ground-based telescope, $k(x,y)$ is the point spread function (PSF) associated with atmospheric turbulence, characterizing the blurring effects \cite{racine1996telescope}. Once the observation $g$ is acquired, the essential information for assessing the sharpness of the image quality is the PSF. 

The PSF, in the absence of atmospheric effects, is determined solely by the telescope aperture \cite{goodman1985statistical}. However, in the presence of atmospheric turbulence, wavefront aberrations distort the PSF, which in turn enables the prediction of light behavior by analyzing the wavefront phase in the pupil space \cite{rousset1999wave}. Following a Fourier optics model \cite{bardsley2008wavefront, goodman2005introduction}, the PSF can be written as a function of the incoming wavefront:
\begin{equation}
    \label{adaptive_optics_system}
    k(x,y) = \left|\mathcal{F}^{-1}\left(W(x,y)e^{\iota \boldsymbol{\phi}(x,y)}\right)\right|^{2},
\end{equation}
where $\mathcal{F}$ denotes the Fourier transform and $\mathcal{F}^{-1}$ represents its inverse, $W(x,y)$ is the indicator function describing the telescope aperture function (taking the value $1$ inside the aperture and $0$ otherwise), $\iota = \sqrt{-1}$ and $\boldsymbol{\phi}(x,y)$ is the phase of the incoming wavefront.

\begin{figure}[htbp]
    \centering
    \includegraphics[width=0.86\textwidth]{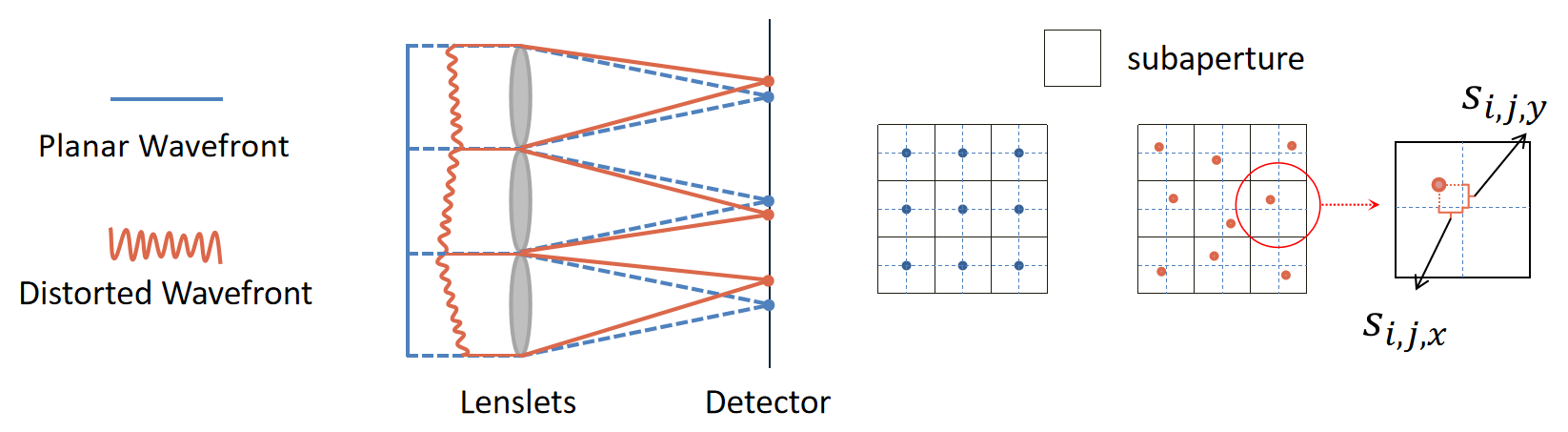}
    \caption{Working principle of Shack Hartman WFS. The offsets $s_{i,j,x}$ and $s_{i,j,y}$ are the horizontal and vertical wavefront gradients detected by the subaperture on the $i$-th row and $j$-th column respectively.}
    \label{fig: WFS_lens}
\end{figure}

\begin{figure}[htbp]
    \centering
    \includegraphics[width=0.62\textwidth]{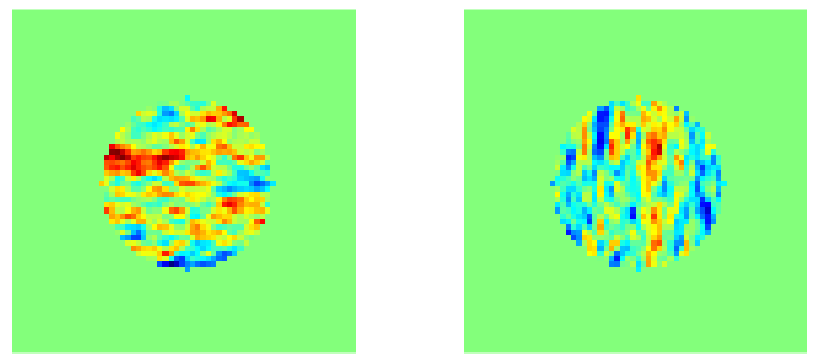}
    \caption{Visualization of WFS observations: the horizontal wavefront gradient (left), and the vertical wavefront gradient (right).}
    \label{fig: Phase_and_grad}
\end{figure}

To obtain the wavefront, modern ground-based telescopes use Adaptive Optics (AO) systems to indirectly obtain $\boldsymbol{\phi}(x,y)$ by the wavefront gradient measured by a wavefront sensor (WFS)\cite{davies2012adaptive, roddier1999adaptive}. The WFS is a critical component engineered to measure the wavefront \cite{wyngaard1992atmospheric}. For our discussion, we assume the implementation of an AO system with a Shack-Hartmann WFS \cite{platt2001history, Shack71}. The Shack-Hartmann WFS consists of multiple lenslets, each focusing on the corresponding part of the wavefront. Every lenslet corresponds to one subaperture and focuses the wavefront beam onto the detector. By analyzing the position of the focus relative to the planar wavefront focus, the offset information of the wavefront in the horizontal and vertical directions is obtained (see Fig. \ref{fig: WFS_lens}), which relates to the $x$ and $y$ components of the WFS. It is the horizontal and vertical wavefront gradient, the measurements of the WFS (see Fig. \ref{fig: Phase_and_grad}).

An AO system is designed to compensate for the distortions of the incoming phase caused by atmospheric turbulence and thereby make the observed images sharper. This correction is typically achieved by adjusting the surface of a so-called deformable mirror (DM). However, due to various physical limitations and operational delays (such as the acquisition of measurements by the WFS and the update of the DM), these distortions caused by atmospheric turbulence can never be entirely corrected. As a consequence, we focus on reconstructing the wavefront phase from wavefront gradients to compensate for the missing information that the AO system can not correct. This process typically incorporates a time-averaged approach, as the exposure time for a single astronomical science image is considerably longer than the frame rate of the WFS.

In this paper, we propose a method to reconstruct wavefronts over multiple consecutive time steps with multiple WFSs at a finer resolution than that provided by the WFS measurements. Rather than targeting real-time adaptive optics control as, e.g., in \cite{Ono_2016, oberti2022super}, our method is formulated as a post-processing framework, enabling the joint exploitation of sequential WFS measurements over multiple time steps. In particular, we reconstruct the \textit{residual wavefront after AO correction}. These reconstructions can then be used to improve quality metrics for the observed images, e.g., through using them as input to methods for PSF reconstruction as in \cite{WaHoRa18, Wagner_2022,WaNiRa23}. In particular, our method could serve as pre-processing step to running a PSF reconstruction pipeline as, e.g., in \cite{MICADO24_psfr_software, MICADO24_psfr_offaxis}. Our approach integrates a sequence of WFS measurements while accounting for atmospheric shifts caused by the wind. Specifically, we model these shifts using a motion matrix. The down-sampling operator between the different grids, along with the telescope aperture mask, results in an ill-posed system of linear equations that must be solved. To ensure a stable solution, we incorporate a penalty term based on the statistical properties of atmospheric turbulence.

The remainder of this paper is structured as follows: In Section~\ref{sec: model}, we model our problem using motion matrices between different WFS frames and taking multiple WFSs into account. We recall some mathematical preliminaries in Section~\ref{sec: H2L2} and formulate the system of linear equations to be solved. We perform different tests and compare our approach to the other existing methods in Section~\ref{sec: simulation}.

\section{Problem modeling}\label{sec: model}
When incorporating multiple WFSs into the model, we simultaneously capture data from multiple regions within each time step, while also considering multiple time steps. In this chapter, we develop our model to utilize all detection information from multiple WFSs across consecutive time steps.

\subsection{WFS model} \label{sec: Macroscopic model}
Several recent developments in super-resolution wavefront reconstruction have leveraged simultaneous multi-WFS measurements, where the different guide-star geometries provide naturally shifted sampling patterns that enable super-resolution within a single time step \cite{fusco2022key, oberti2022super}. Using WFS rotation to achieve a form of \emph{natural super resolution} has also been proposed as an auxiliary reconstruction strategy \cite{cranney2024mavis, cranney2021optimising}. These approaches, while effective, rely on instantaneous multi-view information. In addition, our method exploits temporal diversity by combining measurements over multiple time steps, achieving super-resolution through a multi-time-step post-processing framework.

Let us start with the model setup. The exact number of WFS units on the telescope may vary depending on the instrument configurations and observational requirements. Throughout this paper, we denote the number of WFSs employed by $P$ and use small $p$, $p=1, \ldots P$, to represent the $p$-th WFS. In addition, WFS measurements are taken repeatedly at discrete time steps $t$, $t = 1, 2, \dots, T$, with $T$ representing the total number of time steps. The system operates with a system-dependent time step, $\Delta t$, typically in the millisecond range. 

Due to physical limitations, the subaperture size must be large enough to facilitate adequate light absorption for proper device functionality. This enlargement ultimately leads to a substantially limited resolution. Consequently, in the observation process, the WFS discretizes the continuous wavefront information into subaperture-based units. To represent this mathematically, the WFS is modeled as an operator $\Gamma = (\Gamma_x, \Gamma_y)^{T}: H^{11/6}(\Omega) \to \mathbb{R}^{n^2 \times 2}$ as in \cite{Helin_2013}, which maps the phase onto its discretized subaperture-based $n \times n$ units, i.e.,
\begin{equation}
    \label{gradient operator definition}
    \begin{bmatrix}
        \boldsymbol{s}'_{x} \\
        \boldsymbol{s}'_{y}
    \end{bmatrix}
    =
    \begin{bmatrix}
        \Gamma_{x} \\
        \Gamma_{y}
    \end{bmatrix}
    \boldsymbol{\phi} + \boldsymbol{\eta},
\end{equation}
where $\boldsymbol{s}'_{x}$ and $\boldsymbol{s}'_{y}$ are the measurements of WFS, and $\boldsymbol{\eta}$ models the noise. 

While the continuous framework provides the theoretical basis for wavefront reconstruction, we adopt a high-resolution discrete approach in our simulations to approximate the continuous conditions. We accordingly introduce the operator $R: \mathbb{R}^{nk \times nk} \to \mathbb{R}^{n \times n}$, where $k \in \mathbb{N}$, which acts as a down-sampling operator. The operator $R$ operates on the high-resolution wavefront data to estimate the continuous wavefront function and maps it to its discretized subaperture-based units. Additionally, we use discretized derivative operators $D = [D_{x}, D_{y}]^{T}$: $\mathbb{R}^{n \times n} \to \mathbb{R}^{2 \times n^2}$ to differentiate the high-resolution phase in both horizontal and vertical directions, discretizing the function of $\Gamma$. 

Moreover, the telescope aperture shape determines the extent of the observable region, confining it to the aperture boundary. To account for this, we include an indicator matrix $W$: $\mathbb{R}^{2n \times n} \to \mathbb{R}^{2n \times n}$, which grabs the aperture area. Thus, the actual WFS should be modeled as a compound operator $\mathbb{R}^{nk \times nk} \to \mathbb{R}^{2 \times n^2}$, reflecting the mapping from $\boldsymbol{\phi}$ onto the wavefront gradient, which has a low resolution:
\begin{equation}
    \label{WDR}
    \begin{bmatrix}
        \boldsymbol{s}_{x,t}^{p} \\
        \boldsymbol{s}_{y,t}^{p}
    \end{bmatrix}
    = W
    \begin{bmatrix}
        D_{x} \\
        D_{y}
    \end{bmatrix}
    R\boldsymbol{\phi}^{p}_{t} + \boldsymbol{\eta}_{t}^{p}
    = WDR\boldsymbol{\phi}^{p}_{t} + \boldsymbol{\eta}_{t}^{p},
\end{equation}
where $\boldsymbol{\phi}^{p}_{t}$, $\boldsymbol{s}_{x,t}^{p}$ and $\boldsymbol{s}_{x,t}^{p}$ represent the high-resolution phase and the wavefront gradients in the horizontal and vertical directions, respectively, detected by the $p$-th WFS at time step $t$. Here, $R\boldsymbol{\phi}^{p}_{t}$ can be viewed as a joint representation of the phase on a coarse grid, where each pixel corresponds to a subaperture on the WFS detector. The pixels on the coarse grid are employed to estimate the signals detected by individual subaperture. 

At this point, we have completed the mapping from the high-resolution phase to the WFS gradient detection. We aim to solve the inverse problem, i.e., to reconstruct the high-resolution phase $\boldsymbol{\phi}^{p}_{t}$. Since we want to include multiple WFSs and detection over a period of time in our model, establishing the relationship between all detection points is crucial. Similar approaches have been proposed in the literature~\cite{Ono_2016, oberti2022super}, but they primarily focused on real-time execution rather than multi-time-step post-processing. In the following sections, we will focus on building and analyzing this relationship.

\begin{figure}[htbp]
    \centering
    \includegraphics[width=0.76\textwidth]{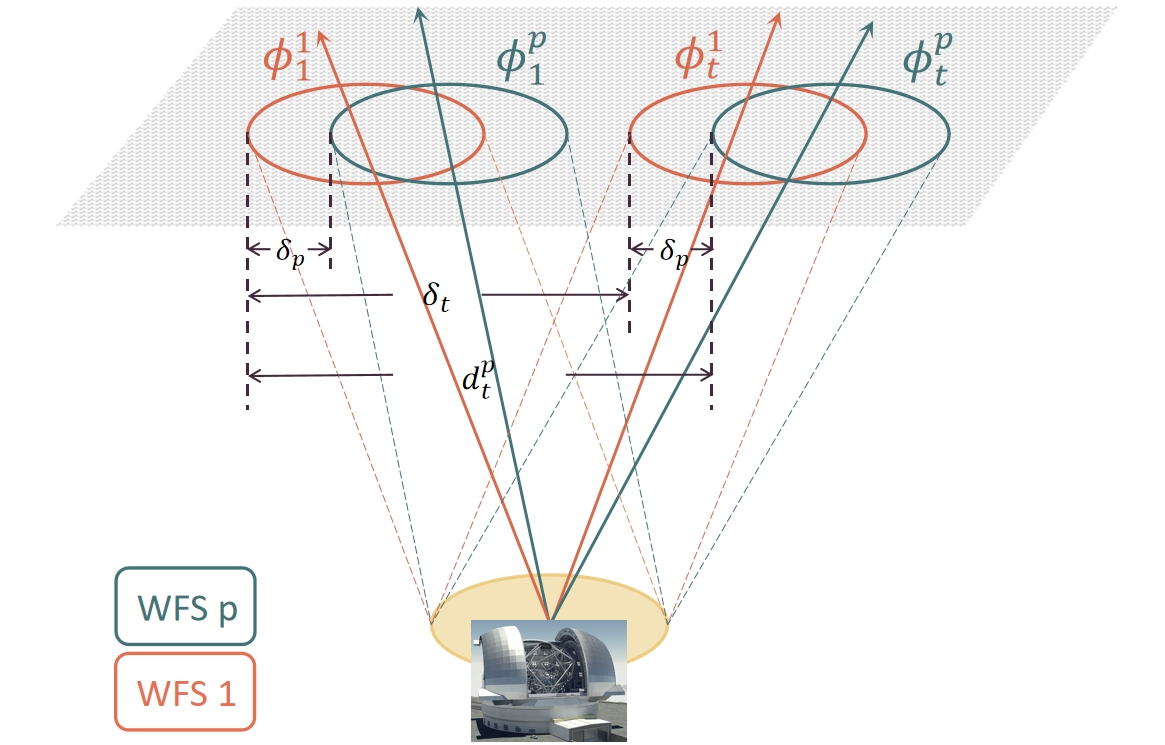}
    \caption{An exaggerated illustration of the multi-WFS model to show the relationship clearly: The orange line represents the information related to WFS $1$, while the green line depicts the information associated with the $p$-th WFS. Here $\boldsymbol{\delta}_{p}$ denotes the inherent distance, i.e., the distance between $\boldsymbol{\phi}_{t}^{1}$ and $\boldsymbol{\phi}_{t}^{p}$, and $\boldsymbol{\delta}_t$ indicates the extrinsic distance, i.e., the distance between $\boldsymbol{\phi}_{1}^{p}$ and $\boldsymbol{\phi}_{t}^{p}$.}
    \label{fig: WFS2_d1_d2}
\end{figure}

\subsection{Multi-WFS model}
Modern telescopes like the {\it Extremely Large Telescope (ELT)} of the European Southern Observatory (ESO) do not only accommodate Single Conjugate Adaptive Optics (SCAO) systems, featuring one wavefront sensor and one deformable mirror, but also Multi Conjugate Adaptive Optics (MCAO) systems containing multiple WFSs and multiple deformable mirrors. The goal of an MCAO system is the reconstruction of the turbulence distribution above the telescope in order to use the deformable mirrors to correct the image over a larger field of view \cite{Fusco, YuHeRa13b, StadlerRamlau2021, StRa22}. E.g., the ELT MCAO system MORFEO features six laser guide stars with associated WFSs and three low-resolution WFSs for faint natural guide stars \cite{10.1117/12.2628969short,capasso2024morfeo}. However, the tomographic reconstruction of the turbulent layers in the system is based on the relatively low resolution of each WFS, also resulting in a low-resolution tomographic reconstruction. Most such systems involve multiple DMs and require real-time solutions, with dedicated methods described, for example, in \cite{StadlerRamlau2021,thiebaut2010fast, tallon2010fractal, tallon2011performances, brunner2012optimal, StRa22, StBiMaRa2021, StBiMaRa19, SaRa15}, where to our knowledge, only FEWHA \cite{StadlerRamlau2021} meets the stringent time requirements. In contrast, we consider a special case of MCAO with a single DM conjugated to the ground layer and employ multiple WFSs in multiple time steps as a post-processing. Our goal is to achieve a high-resolution wavefront at the center of the telescope, even though the DM resolution itself remains relatively coarse.

Recall that measurements for multiple time steps are considered, and the atmospheric turbulence distribution varies over time. In the multi-WFS model, the WFSs are located at different positions throughout the observation process but maintain a fixed detection direction close to the target object. During the observation, a continuous evolution of turbulence occurs in the atmospheric layers due to the interplay between solar heating, temperature differences, the Coriolis effect, and pressure variations \cite{davidson2015turbulence, roddier1981v}. Consequently, each WFS obtains distinct wavefront gradients in the observed area at various time steps due to the atmospheric evolution.

To model multiple WFSs, we select a reference frame and establish the positional relationships among all the frames. We begin our model by introducing $\boldsymbol{\phi}_{1}^{1}$. In Fig.~\ref{fig: WFS2_d1_d2}, an exaggerated illustration of a two-WFS case across two separate time steps is depicted. The representation introduces two crucial distances: the inherent distance denoted as $\boldsymbol{\delta}_{p}$ and the extrinsic distance denoted as $\boldsymbol{\delta}_t$. The inherent distance signifies the effect of the detection positions of the WFSs, and the extrinsic distance represents the impact of the movement of atmospheric turbulence on the current turbulence layer. Specifically, $\boldsymbol{\delta}_p$ reflects the inherent distance between $\boldsymbol{\phi}_{t}^{1}$ and $\boldsymbol{\phi}_{t}^{p}$, signifying the separation between frames detected by the first WFS and the $p$-th WFS at the same time step $t$. While $\boldsymbol{\delta}_t$ represents the extrinsic distance between $\boldsymbol{\phi}_{1}^{p}$ and $\boldsymbol{\phi}_{t}^{p}$, explaining the movement of the layer from time step $1$ to time step $t$.

Consequently, the overall displacement $\boldsymbol{d}_{t}^{p}$ between $\boldsymbol{\phi}_{t}^{p}$ and the reference frame $\boldsymbol{\phi}_{1}^{1}$ can be characterized as:
\begin{equation}
    \label{eq: dpt}
    \boldsymbol{d}_{t}^{p} = \boldsymbol{\delta}_{p} + \boldsymbol{\delta}_t,
\end{equation}
where $\boldsymbol{\delta}_{p}$ is a known fixed displacement between the first WFS and the $p$-th WFS. Specifically, to clarify the relationships among all frames, we introduce the full phase $\boldsymbol{\Phi}_{t}^{p}$, which corresponds to the full phase across the entire detection area. The precise definition is given by $\boldsymbol{\phi}_t^p = W \boldsymbol{\Phi}_t^p$, where $\boldsymbol{\phi}_t^p$ captures only the information available through the aperture. This full-phase information is used to estimate the wavefront by referring to the mask-covered information. Due to the continuously shifting turbulence affecting different detection areas, a full-range estimation is required to capture these relationships. Thus, the displacement relationship in the full-range estimation can be described by the following formula:
\begin{equation}
    \label{eq: dp}
    \boldsymbol{\Phi}_{t}^{p}(x) = \boldsymbol{\Phi}_{t}^{1}(x+\boldsymbol{\delta}_{p}).
\end{equation}

\subsection{Turbulence Model}
To model $\boldsymbol{\delta}_t$, we incorporate the wind motion, introduced by the atmospheric model to simulate the dynamic changes of the atmosphere \cite{tennekes1972first}. More specifically, Taylor's frozen flow hypothesis (FFH), a fundamental concept in the field of turbulent flows \cite{taylor1938spectrum} is introduced. The FFH assumes that within a turbulent flow, small fluid elements or particles are effectively ``frozen'' into the local velocity field over short time intervals \cite{nagy2010fast}. In other words, this turbulence preserves its velocity and relative position during our observation \cite{bharmal2015frozen}. Each time step takes only a few milliseconds of real-time in our application. Therefore, the FFH hypothesis enables the connection of turbulence information across consecutive time steps. 
The incoming phase reaching the telescope has a translation as
\begin{equation}
    \label{motion_relationship}
    \boldsymbol{\Phi}_{t}^{p}(x) = \boldsymbol{\Phi}_{1}^{p}(x-(t-1)\boldsymbol{v}) = \boldsymbol{\Phi}_{1}^{p}(x-\boldsymbol{\delta}_t)
\end{equation}
where $\boldsymbol{v}$ represents the wind velocity. Wind speeds can be estimated from prior measurements or inferred from models that adapt wind speeds during reconstruction \cite{ke2020reconstruction}. Here, we adopt a known wind speed to focus on the reconstruction framework.

\begin{figure}[htbp]
    \centering
    \includegraphics[width=0.86\textwidth]{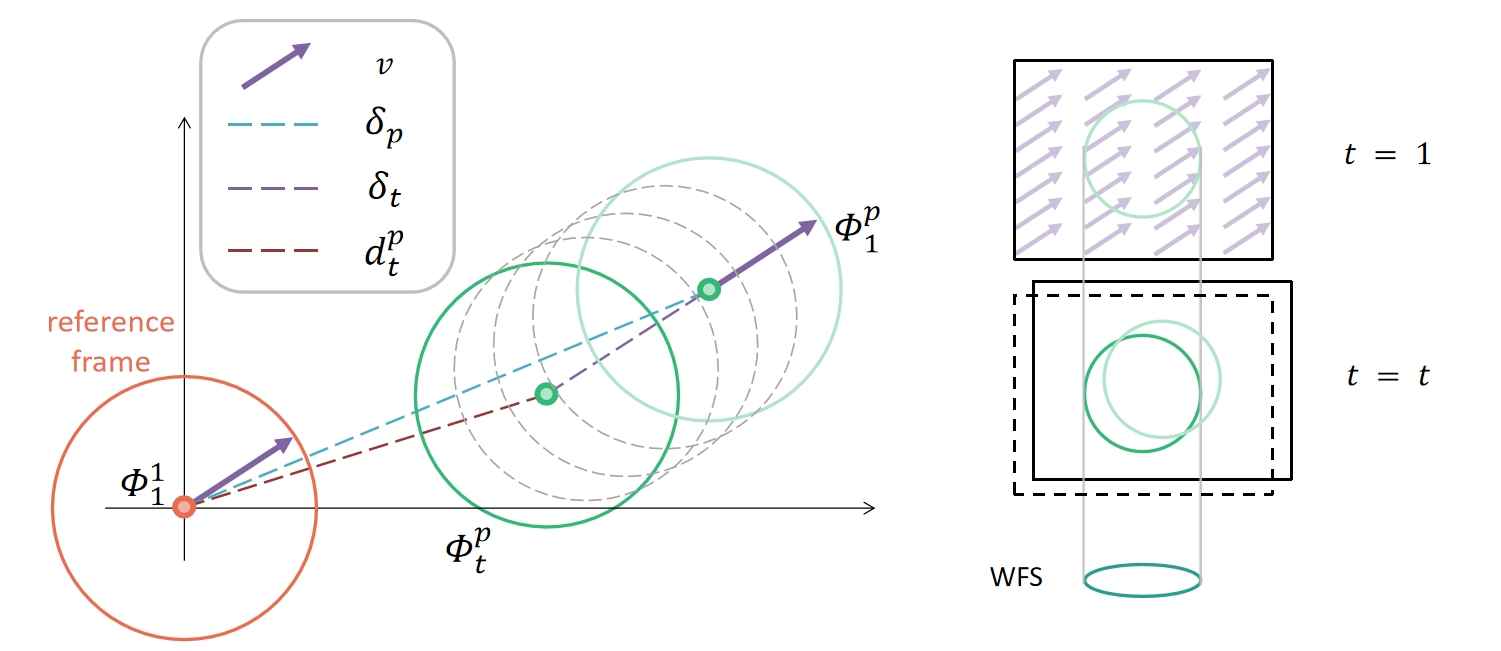}
    \caption{The positional relationship from a top-view perspective. The right figure illustrates the atmospheric motion influenced by wind relative to the ground, emphasizing the detection areas at different time steps under the fixed ground-based telescope. The left figure provides a detailed explanation of the positional relationship between $\boldsymbol{\phi}_{t}^{p}$ and $\boldsymbol{\phi}_{1}^{p}$ as shown in the right figure, highlighting the influence of wind. The inherent distance $\boldsymbol{\delta}_{p}$, the extrinsic distance $\boldsymbol{\delta}_{t}$, and the overall distance $\boldsymbol{d}_{t}^{p}$ are depicted using light blue, red, and purple dotted lines, respectively.}
    \label{fig: Top_view}
\end{figure}

We recognize that the extrinsic distance  $\boldsymbol{\delta}_t = -(t-1)\boldsymbol{v}$ is the displacement of turbulence from time step $1$ to time step $t$, and it indicates the cumulative effect of wind over the given $t-1$ time steps. Subsequently, according to \eqref{eq: dpt}, \eqref{eq: dp} and \eqref{motion_relationship}, $\boldsymbol{\Phi}_{t}^{p}$ with reference to $\boldsymbol{\Phi}_{1}^{1}$ can be expressed by the following equation via the displacement $\boldsymbol{d}_{t}^{p}$:
\begin{equation}
    \label{dtp_relationship}
    \boldsymbol{\Phi}_{t}^{p}(x) = \boldsymbol{\Phi}_{1}^{1}(x+\boldsymbol{d}_{t}^{p}), \quad \boldsymbol{d}_{t}^{p} = \boldsymbol{\delta}_{p} - (t-1)\boldsymbol{v}.    
\end{equation}
The inherent and extrinsic distances represented by the introduction of wind speed are illustrated in the top view shown in Fig. \ref{fig: Top_view}. 

\begin{figure}[htbp]
    \centering
    \includegraphics[width=0.8\textwidth]{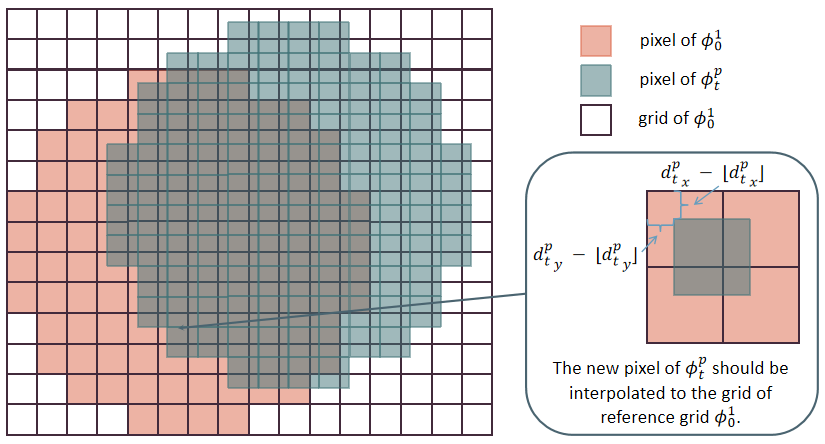}
    \caption{The grid of $\boldsymbol{\phi}_{1}^{1}$ is shifted by $\boldsymbol{d}_{t}^{p}$ to $\boldsymbol{\phi}_{t}^{p}$, resulting in a configuration where it no longer aligns precisely with the grid (i.e., involving fractional movement). To delineate this fractional movement, the displaced field $\boldsymbol{\phi}_{t}^{p}$ is interpolated onto the grid of $\boldsymbol{\phi}_{1}^{1}$ by integral movement $\lfloor {\boldsymbol{d}_{t}^{p}}_x \rfloor$ and $\lfloor {\boldsymbol{d}_{t}^{p}}_y \rfloor$, and fractional movement ${\boldsymbol{d}_{t}^{p}}_x - \lfloor {\boldsymbol{d}_{t}^{p}}_x \rfloor$ and ${\boldsymbol{d}_{t}^{p}}_y - \lfloor {\boldsymbol{d}_{t}^{p}}_y \rfloor$. Here, ${\boldsymbol{d}{t}^{p}}x$ and ${\boldsymbol{d}{t}^{p}}y$ denote the horizontal and vertical components of the displacement $\boldsymbol{d}_{t}^{p}$, respectively.}
    \label{fig: pixel}
\end{figure}

To elucidate the relationship, we introduce the discrete motion operator $A_{t}^{p}$ to describe the interpolated shifts, $\boldsymbol{d}_{t}^{p}$. The fractional displacement is estimated using bilinear interpolation, which linearly weights the four nearest pixels according to the fractional displacement. Figure~\ref{fig: pixel} gives a reference illustration for estimating the interpolation weights corresponding to the fractional displacement relative to the four neighboring pixels. Consequently, \eqref{dtp_relationship} can be approximated discretely as:
\begin{equation}
    \label{Atp_operator}
    A_{t}^{p}\boldsymbol{\Phi}_{1}^{1} = \boldsymbol{\Phi}_{t}^{p}.
\end{equation}
Together with with \eqref{Atp_operator} and \eqref{WDR}, we give the formulation of the observation model as
\begin{equation}
    \label{WDRA}
    \boldsymbol{s}_{t}^{p} = WDR\boldsymbol{\Phi}_{t}^{p} + \boldsymbol{\eta}_{t}^{p} = WDRA_{t}^{p}\boldsymbol{\Phi}_{1}^{1} + \boldsymbol{\eta}_{t}^{p},
\end{equation}
where $\boldsymbol{s}_{t}^{p}:= [\boldsymbol{s}_{x,t}^{p}, \boldsymbol{s}_{y,t}^{p}]^{T}$ are the captured low-resolution wavefront gradients on the coarse grid at time step $t$ by the $p$-th WFS. 

Problem \eqref{WDRA} is similar to that of multi-frame super-resolution in the realm of image reconstruction, see e.g., \cite{bose1998high, tsai1984multiframe, chan2003wavelet}. In summary, we adopt the following standard variational approach by adding a regularization term:
\begin{equation}
    \label{minimization}
    \underset{\boldsymbol{\Phi}}{\min} \frac{1}{2} \sum_{p = 1}^{P} \sum_{t = 1}^{T} ||WDRA_{t}^{p} \boldsymbol{\Phi} - \boldsymbol{s}_{t}^{p}||_{2}^{2}  + \alpha||\boldsymbol{\Phi}|| ,
\end{equation}
where $\boldsymbol{\Phi}$ represents the discrete phase function, where each value corresponds to a point-based approximation of the underlying continuous phase distribution. It represents an arbitrary reference frame, indicating that any frame could serve as the reference without a specific designation. For the term that remains unspecified, we want to find an appropriate space, especially for atmospheric images, and thus an appropriate norm $||\cdot||$ for the penalty term using known properties of the atmospheric turbulence.

\section{$L^2$-$H^2$ Model} \label{sec: H2L2} 
\subsection{Overview of Existing Methods}
Regarding the penalty term, the $L^1$ norm is well known for its ability to sharpen edges in solutions, see, e.g., \cite{chan2005image}. The $L^2$-$L^1$ model was subsequently proposed in \cite{chan2013phase}. By the notation $L^2$-$L^1$, we refer to the combination of norms for the Residual and the penalty term. Incorporating an $L^1$ regularization term for multi-WFS reconstruction (\ref{minimization}) leads to the following formulation:

\begin{equation}
    \label{eq: l1l2}
    \underset{\boldsymbol{\Phi}}{\min} \frac{1}{2} \sum_{p = 1}^{P} \sum_{t = 1}^{T} ||WDRA_{t}^{p} \boldsymbol{\Phi} - \boldsymbol{s}_{t}^{p}||_{2}^{2}  + \tau_1 ||C\boldsymbol{\Phi}||_1 ,
\end{equation}
where $C$ is a tight frame satisfying the orthogonality condition $C^{T}C = I$. Subsequently, an $L^2$-$L^2$ model utilizing $L^2$ regularization was proposed \cite{ke2020reconstruction}. When applied to multi-WFS reconstruction, the resulting optimization problem is given as follows:
\begin{equation}
    \label{eq: l2l2}
    \underset{\boldsymbol{\Phi}}{\min} \frac{1}{2} \sum_{p = 1}^{P} \sum_{t = 1}^{T} ||WDRA_{t}^{p} \boldsymbol{\Phi} - \boldsymbol{s}_{t}^{p}||_{2}^{2}  + \tau_2 ||H_1\boldsymbol{\Phi}||_2^2 ,
\end{equation}
where $H_1^2$ approximates the inverse covariance matrix of $\boldsymbol{\phi}$ up to a scaling constant, according to the statistics of the turbulence. Besides that, the joint regularization method denoted as $L^2$-$L^2L^1$, which combines $L^1$ and $L^2$ regularization terms, has also been shown to achieve good performance in image reconstruction \cite{zou2005regularization, cai2013two}. It yields the following minimization problem:
\begin{equation}
    \label{eq: l1l2l2}
    \underset{\boldsymbol{\Phi}}{\min} \frac{1}{2} \sum_{p = 1}^{P} \sum_{t = 1}^{T} ||WDRA_{t}^{p} \boldsymbol{\Phi} - \boldsymbol{s}_{t}^{p}||_{2}^{2}  + \tau_3||H_1\boldsymbol{\Phi}||_2^2 + \tau_4||C\boldsymbol{\Phi}||_1.
\end{equation}

While these existing models have produced satisfactory results, their optimization constraints do not align with the inherent characteristics of atmospheric turbulence. Therefore, we consider incorporating a regularization term that accounts for statistical properties, computational complexity, and discretization errors in the observations.

\subsection{Turbulence Statistics}
Turbulence statistics are usually described by the power spectral density (PSD). In this paper, we assume the von Karman PSD of the phase $\boldsymbol{\phi}$ (see, e.g., \cite{von1931mechanical, roggemann2018imaging}), which is given by
\begin{equation}
    \label{Von_karman}
    \Psi_n(\kappa) = \frac{0.033C_{n}^{2}}{(\kappa^2 + \kappa_{0}^{2})^{11/6}} \exp\left(-\frac{\kappa^2}{\kappa_{m}^{2}}\right),
\end{equation}
where $\kappa$ is the spatial frequency of the turbulence, $C_{n}^{2}$ is the local strength of the index of refraction fluctuation \cite{hill1978spectra}, $\kappa_m = 5.92/l_0$ and $\kappa_0 = 2\pi/L_0$. $l_0$ and $L_0$ are the inner scale and outer scale respectively \cite{hill1992review}.

Explicitly, this power spectrum exhibits a power-law decay of $\kappa^{-11/6}$. This power-law exponent of $11/6$ precisely matches the regularity requirement of the $H^{11/6}$ Sobolev space \cite{tatarskii1971effects, Helin_2013}, illustrating the inherent regularity of $\boldsymbol{\phi}$ affected by atmospheric turbulence.

\subsection{Approximating $H^{11/6}$ by $H^{2}$}
In the following, we will use a Sobolev space to characterize the turbulence statistics and find a suitable regularization term. Specifically, we use the previously introduced von Karman power law \eqref{Von_karman} to set up a penalty term in a Sobolev space. For an introduction to Sobolev spaces, we refer to \cite{adams2003sobolev, di2012hitchhikerʼs}.

The Sobolev space $H^{11/6}$ is a Banach space with respect to the norm
\begin{equation}
    ||u||_{H^{11/6}(\Omega)} = \left(||u||_{2}^{2} + ||\nabla u||_{2}^{2} + ||\nabla u||_{H^{5/6}(\Omega)}^{2}\right)^{\frac{1}{2}},
\end{equation}
where
\begin{equation}
    ||u||_{H^{5/6}(\Omega)} := \left(  \int_{\Omega} \frac{|u(x)-u(y)|^{2}}{|x-y|^{n+ \frac{5}{3}}} d(x,y) \right)^{\frac{1}{2}},
\end{equation}
and $n$ is the dimension, i.e., $n=2$ in our application.

In our study, we recognize that the computational complexity introduced by the $H^{11/6}$-regularization significantly increases the cost of solving optimization problems, potentially exceeding the benefits in terms of accuracy. Furthermore, minimizing the impact of discretization noise during detection is a critical component of our reconstruction strategy. Motivated by these considerations, we want to use a higher-order norm, namely one for the space $H^{2}$.

From \cite{di2012hitchhikerʼs}, we directly obtain the embedding 
\begin{equation}
    \label{H_m+1_s}
    H^{2}(\Omega) \subseteq H^{11/6}(\Omega).
\end{equation}
In other words, choosing $H^2$ as the regularization term gives results that are also in $H^{11/6}$. Although the $H^2$-penalty term enforces a smoother phase than turbulence statistics suggest, this approach also mitigates the discretization noise inherent in atmospheric modeling. The $H^{2}$-norm is defined as:
\begin{equation}
    \label{H2_norm}
    \lVert u \rVert_{H^{2}} = \lVert u \rVert_{W^{m,2}} = (\lVert u \rVert_{2}^{2} + \lVert \nabla u \rVert_{2}^{2} + \lVert \nabla^2 u \rVert_{2}^{2})^{1/2}.
\end{equation}
Consequently, it significantly reduces computational demands. In essence, this represents a pragmatic trade-off between capturing the essential statistical features of the phase and maintaining computational efficiency.

Given the Fourier optics model \eqref{adaptive_optics_system}, we observe that the magnitude of the phase does not influence the PSF. Additionally, the WFS detects wavefront gradients and does not yield information about the magnitude of turbulence. In other words, the solution space of \eqref{minimization} is expressed as $\boldsymbol{\Phi} + a\bm{1}_{kn \times kn}$, where $a \in \mathbb{R}$, i.e., we are not able to determine constants in the reconstruction, like the constant term of the original function in the indefinite integral problem \cite{rudin1964principles}. In our model, we ensure that our solution has zero mean and therefore, we remove the term $||\boldsymbol{\Phi}||_{2}^{2}$. In the discrete setting, we approximate the gradient $\nabla$ using finite difference operators. Specifically, for a discrete phase field $\boldsymbol{\Phi} \in \mathbb{R}^{kn \times kn}$, we define the gradient operator as  
\begin{equation}
    \nabla \boldsymbol{\Phi} = G \boldsymbol{\Phi} =
    \begin{bmatrix}
        G_x \boldsymbol{\Phi} \\ 
        G_y \boldsymbol{\Phi}
    \end{bmatrix},
\end{equation}
where $G_x$ and $G_y$ are the finite difference matrices in the $x$- and $y$-directions, respectively. These operators are constructed using the Kronecker product as follows:
\begin{equation}
    G_x = I_{kn} \otimes G_1, \quad G_y = G_1 \otimes I_{kn},
\end{equation}
where $G_1 \in \mathbb{R}^{kn \times kn}$ is the one-dimensional forward difference matrix given by
\begin{equation}
    G_1 = 
    \begin{bmatrix}
        -1 & 1 & 0 & \cdots & 0 \\
        0 & -1 & 1 & \cdots & 0 \\
        \vdots & \vdots & \vdots & \ddots & \vdots \\
        0 & 0 & 0 & \cdots & -1
    \end{bmatrix} \in \mathbb{R}^{kn \times kn}.
\end{equation}
Additionally, we consider $\nabla^2 \approx L$, where $L$ is the discrete Laplacian operator. The discrete Laplacian matrix is defined as 
\begin{equation}
    L = I_{kn} \otimes L_1 + L_1 \otimes I_{kn},
\end{equation}
where $L_1 \in \mathbb{R}^{kn \times kn}$ is the one-dimensional discrete Laplacian matrix given by
\begin{equation}
    L_1 =
    \begin{bmatrix}
        -2 & 1 & 0 & \cdots & 0 \\
        1 & -2 & 1 & \cdots & 0 \\
        0 & 1 & -2 & \cdots & 0 \\
        \vdots & \vdots & \vdots & \ddots & \vdots \\
        0 & 0 & 0 & \cdots & -2
    \end{bmatrix} \in \mathbb{R}^{kn \times kn}.
\end{equation}
Thus, we approximate the Laplacian of $\boldsymbol{\Phi}$ as  
\begin{equation}
    \nabla^2 \boldsymbol{\Phi} \approx L \boldsymbol{\Phi}.
\end{equation}
For the convenience of notation, let
\begin{equation}
    ||u||_{\Tilde{H}^2}^2 = ||\nabla u||_2^2 + ||\nabla^2 u||_2^2 = ||G u||_2^2 + ||L u||_2^2,
\end{equation}

Based on \eqref{minimization}, we propose the following function for the phase reconstruction problem
\begin{equation}
    \label{argmin_H2}
    \underset{\boldsymbol{\Phi}}{\min} \frac{1}{2} \sum_{p = 1}^{P} \sum_{t = 1}^{T} ||WDRA_{t}^{p} \boldsymbol{\Phi} - \boldsymbol{s}_{t}^{p}||_{2}^{2}  + \alpha(||G\boldsymbol{\Phi}||_2^2 + ||L\boldsymbol{\Phi}||_2^2), 
\end{equation}
where $\int_{\Omega}\boldsymbol{\Phi} \, dx = 0$. The minimizer is given by solving the normal equation. 

\subsection{Numerical Minimization}
In this section, we give the solution to \eqref{argmin_H2}. Our model aligns with the principles of Tikhonov regularization by incorporating multiple regularization terms \cite{tikhonov1943stability}. In the general form of Tikhonov regularization, we can incorporate multiple regularization terms, each with its own penalty parameter. The objective function becomes:
\begin{equation}
    \min_x  \sum_{k=1}^K \|B_k x - b_k\|_2^2 + \lambda_1 \|C_1 x\|_2^2 + \lambda_2 \|C_2 x\|_2^2 ,
\end{equation}
where $B_k \in \mathbb{R}^{m_k \times n}$ are the data matrices for $k = 1, \dots, K$, $C_1$ and $C_2$ are the regularization operators, and $\lambda_1, \lambda_2$ are penalty parameters.

The solution can be obtained by minimizing the objective function. Taking the gradient with respect to $x$ and setting it to zero gives the normal equations:
\begin{equation}
    \sum_{k=1}^K B_k^T B_k x + \lambda_1 C_1^T C_1 x + \lambda_2 C_2^T C_2 x = \sum_{k=1}^K B_k^T b_k
\end{equation}
Rearranging, we get the solution:

\begin{equation}
    x^* = \left( \sum_{k=1}^K B_k^T B_k + \lambda_1 C_1^T C_1 + \lambda_2 C_2^T C_2 \right)^{-1} \left( \sum_{k=1}^K B_k^T b_k \right).
\end{equation}

For our phase reconstruction problem, these matrices correspond to: $B_k = WDRA_t^p$, $C_1 = G$, $C_2 = L$. The objective function becomes:
\begin{equation}
    \min_{\boldsymbol{\Phi}} \sum_{p = 1}^{P} \sum_{t = 1}^{T} \left\| WDRA_{t}^{p} \boldsymbol{\Phi} - \boldsymbol{s}_{t}^{p} \right\|_2^2 
    + \alpha (\left\| G \boldsymbol{\Phi} \right\|_2^2 
    + \left\| L \boldsymbol{\Phi} \right\|_2^2).
\end{equation}
Thus, the normal equation for our problem is given by:
\begin{equation} \sum_{p = 1}^{P} \sum_{t = 1}^{T} (WDRA_{t}^{p})^T (WDRA_{t}^{p}) \boldsymbol{\Phi} + \alpha G^T G \boldsymbol{\Phi} + \alpha L^T L \boldsymbol{\Phi} = \sum_{p = 1}^{P} \sum_{t = 1}^{T} (WDRA_{t}^{p})^T \boldsymbol{s}_{t}^{p}. \end{equation}
Rearranging, we obtain the solution:
\begin{equation}
\begin{split}
    \boldsymbol{\Phi}^* = & \left( \sum_{p = 1}^{P} \sum_{t = 1}^{T} (WDRA_{t}^{p})^T (WDRA_{t}^{p}) + \alpha G^T G + \alpha L^T L \right)^{-1} \\
    & \times \left( \sum_{p = 1}^{P} \sum_{t = 1}^{T} (WDRA_{t}^{p})^T \boldsymbol{s}_{t}^{p} \right).
\end{split}
\end{equation}

In our practical computation, the solution of the linear system is calculated using the conjugate gradient (CG) method, which is well suited for large-scale problems with symmetric positive-definite structure. 

Based on the definitions of the involved operators, $\#(W) \approx 1/n^2$, $\#(D) \approx 2/n^2$, $\#(R) \approx \#(A) \approx 1/n^2$, and $\#(L) \approx \#(G) \approx 2/n^2$, where $\#(\cdot)$ denotes the fraction of non-zero entries per row. Consequently, all operators are highly sparse.
This sparsity significantly reduces memory usage and ensures that each CG iteration can be performed efficiently, resulting in a computationally fast and scalable solver for our problem size.

\subsection{Parameter Selection}
\begin{figure}[h]
\centering
\includegraphics[width=0.7\textwidth]{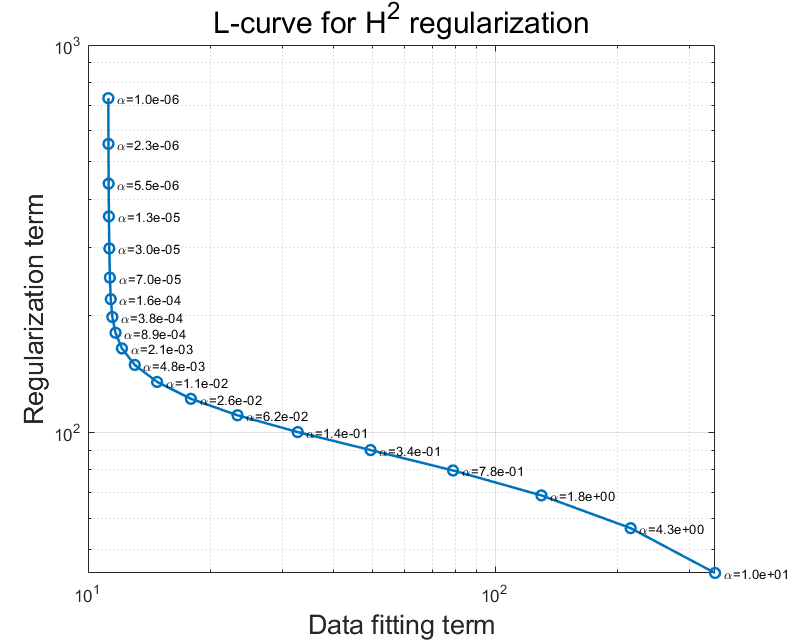}
\vspace{-2mm}
\caption{L-curve used to select the regularization strength $\alpha$.}
\label{fig: L_curve}
\end{figure}

To complete the numerical formulation, we specify the regularization parameter to optimize the trade-off between fitting the WFS measurements and turbulence smoothness. We determine the regularization strength using the data-driven L-curve criterion for our specific task of turbulence reconstruction. Figure~\ref{fig: L_curve} presents the L-curve, which balances the data fitting term, $\|WDRA_{t}^{p} \boldsymbol{\Phi} - \boldsymbol{s}_{t}^{p} \|_2^2$, and the regularization term, $||\boldsymbol{\Phi}||_{\Tilde{H}^2}^2$. The corner region of the L-curve identifies the optimal trade-off between data fidelity and regularization. Within this region, we selected a value $\alpha = 10^{-4}$, slightly toward the data-fitting side, placing more emphasis on preserving fine signal details. Since the $H^2$ regularization is inherently stronger than the theoretically motivated $H^{11/6}$ space according to the statistical properties of the von Karman power law, this choice also mitigates the risk of oversmoothing. This ensures that the $H^2$ regularization effectively suppresses noise while retaining physically meaningful phase features.

\section{Simulation results} \label{sec: simulation}

\subsection{Evaluation Metric} \label{sec: c3EM}
\begin{figure}[htbp]
    \centering
    \includegraphics[width=0.72\textwidth]{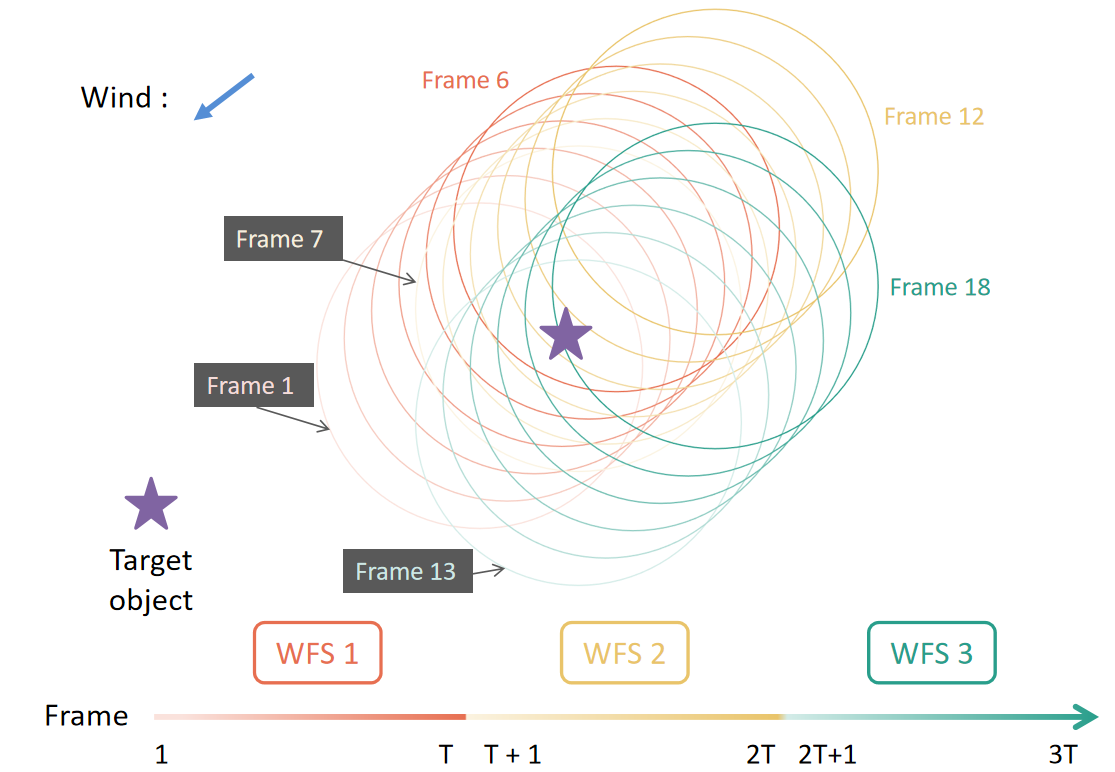}
    \caption{Field of view and frame index. To ensure the image's applicability under various experiment conditions, we use an example case with $3$ WFSs and $6$ time steps. The bottom axis indicates the ordering of the frame index. Each WFS is represented by several gradient lines from light to dark, depicting progression within each time step. The frames are sorted primarily by time step and secondarily by WFS index, with different colors assigned to each WFS.}
    \label{fig: Frame_vis}
\end{figure}

For the simplicity of notation in the visualization of numerical results, we replace $t$ and $p$ by frame index $i$, which we define as $i = (p-1)T + t$. Explicitly, the first $T$ frames correspond to the detection by WFS $1$, from time step $1$ to $T$. Subsequent $T$ frames correspond to the detection by WFS $2$, with $t$ from $1$ to $T$. The remaining frames follow this pattern. Figure~\ref{fig: Frame_vis} presents an exaggerated visual representation of every frame, highlighting the concept of the frame index. Therefore, we replace $\boldsymbol{\phi}_{t}^{p}$ by $\boldsymbol{\phi}_i$, which denotes the wavefront phase at the $i$-th frame, where $i = 1, 2, \dots, M$, $M = T \cdot P$. 

In the following, we introduce two main metrics used to assess the reconstruction performance.\\

\noindent\textbf{Phase Residual Error:} To have a intuitive comparison, the phase residual error $\boldsymbol{\phi}_i^{\mathrm{err}}$ is introduced to display the difference between the reconstruction $\hat{\boldsymbol{\phi}}_i$ and ground truth $\boldsymbol{\phi}_{i}$, which is defined as
\begin{equation}
    \boldsymbol{\phi}_i^{\mathrm{err}} = \hat{\boldsymbol{\phi}_{i}} - \boldsymbol{\phi}_{i}, \quad i = 1, 2, \dots, M,
\end{equation}
where $\hat{\boldsymbol{\phi}_{i}}$ is the corresponding reconstructed phase at the $i$-th frame, i.e., 
\begin{equation}
    \hat{\boldsymbol{\phi}_{i}} = W_{\text{HR}}A_{t}^{p}\hat{\boldsymbol{\Phi}}, \quad  i = (p-1)T + t, \quad i = 1, 2, \dots, M.
\end{equation} 
where $W_{\text{HR}}$ restricts the telescope aperture on fine grid data.
This metric is presented by a figure in our analysis. \\

\noindent\textbf{$\bm{L}_2$ Relative Error and $\bm{H}_1$ Relative Error:} To assess the accuracy of the reconstructed phase $\hat{\boldsymbol{\Phi}}$, we use the $L_2$ relative error and the $H_1$ relative error. Specifically, $L_2$ relative error primarily focuses on the discrepancy between the reconstruction and ground truth, the $H_1$ relative error balances the consideration of both wavefront difference and wavefront gradient difference. Mathematically, they are defined as
\begin{align}
    err_{L_2}(i) &= \frac{\| \boldsymbol{\phi}_i^{\mathrm{err}} \|_2}{\|\boldsymbol{\phi}_{i}\|_2}, \quad i = 1, 2, \dots, M, \\
   err_{H_1}(i) &= \frac{\|\boldsymbol{\phi}_i^{\mathrm{err}}\|_{H_1}}{\|\boldsymbol{\phi}_{i}\|_{H_1}}, \quad i = 1, 2, \dots, M.
\end{align}

\subsection{Experimental Settings} \label{c3ES}
To validate our model, we use a Matlab-based AO simulation to obtain the data. The simulated system is equipped with $3$ WFSs on an $8$ meter telescope and takes into account a dataset comprising $18$ frames of experimental observations across $6$ distinct time steps. The high-resolution incoming phase is estimated by simulation on a fine level with $256 \times 256$ pixels across the telescope aperture. The Shack-Hartmann WFS has $64 \times 64$ subapertures. The noise level is $5\%$, added in the wavefront gradient simulation. We use a seeing parameter $r_0= 0.16~m$ at wavelength $\lambda = 0.744~\mu m$ \cite{goodman2005introduction}. We initiate our simulation with a simplified atmospheric model featuring a single ground layer with a known wind speed of $15 \ m/s$ in the $0^\circ$ direction, employing WFSs in an equilateral triangle arrangement. The fidelity term is established with reference to the third frame.

We compare our model with three established methods from Eqs. \eqref{eq: l2l2}, \eqref{eq: l1l2}, and \eqref{eq: l1l2l2}. As a baseline, we also consider upsampled bilinear interpolation of the wavefront phase based on the gradient operator in \eqref{gradient operator definition}, which uses only a single frame of information.

To demonstrate the effectiveness of our model, we devise a comprehensive evaluation strategy for our model, detailed as follows: (i) We evaluate the performance of different models for one specific dataset (see Section~\ref{sec:Numerical_Results}); Subsequently, we conduct two comparative experiments to investigate the consistency of the model under diverse configurations (see Section~\ref{c3Consistency}); (ii) with fixed WFS positions to examine average performance across different wind directions and speeds, and (iii) with constant wind velocities to analyze the model's performance with varying WFS positions; (iv) Then we focus on our $H^2$ regularization term, we compare with different reference frames for phase reconstruction to search a suitable modeling reference system (see Section \ref{c3Refframe}); (v) Finally, we conduct a comparative analysis between single-WFS and multi-WFS configurations, to elucidate the advantages of multi-WFS in enhancing performance and accuracy (see Section~\ref{c3singlemulti}).

\subsection{Numerical Results} \label{sec:Numerical_Results}
In this section, we present the numerical results for comparing the performance between models. In this simulation, $0^\circ$ wind direction represents an extreme case with an absolute advantage for information in one direction over the other. In other words, one direction may result in a substantial loss of information, rendering it a relatively challenging wind case among all possible directions. The optimal scenario would involve a wind direction at $45^\circ$ to the axis, ensuring an equitable distribution of information gain across both directions. However, such an ideal direction will not always occur in practice.

In particular, according to the L-curve analysis, we choose $\alpha = 10^{-4}$ in our model \eqref{argmin_H2}. For the $L^2$-$L^1$ model and $L^2$-$L^2$ model, we follow the parameters provided in the literature \cite{chan2013phase} and \cite{ke2020reconstruction}, respectively, that is $\tau_1 = 10^{-4}$, and $\tau_2 = 10^{-3}$. For the joint $L^2$-$L^2L^1$ we take $\tau_3 = 10^{-3}$ for the $L_1$ term and $\tau_4 = 10^{-4}$ for the $L_2$ term. The system was implemented via MATLAB’s pcg routine without preconditioning, with a stopping criterion of $1\mathrm{e}^{-8}$ and a maximum of 1000 iterations.

\begin{figure}[htbp]
    \center
    \includegraphics[width=0.92\textwidth]{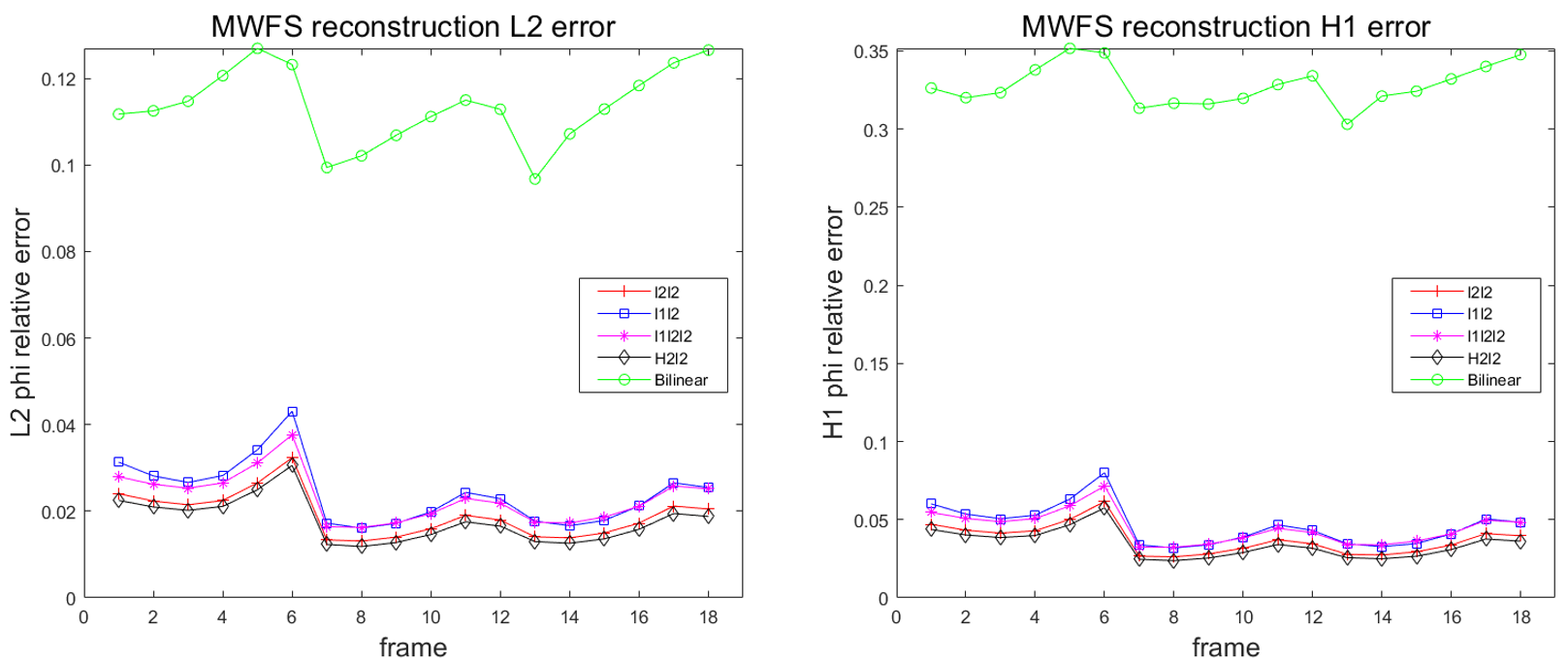}
    \caption{The $L_2$ relative error and $H_1$ relative error of the reconstruction.}
    \label{fig: Bilinear_single}
\end{figure}

In Figure~\ref{fig: Bilinear_single}, we present a visualization plotting the $L_2$ relative error (left) and $H_1$ relative error (right) of the estimated phase $\boldsymbol{\phi}$, with the horizontal axis showing the frame index as described in Section~\ref{sec: c3EM} and the vertical axis depicting the range of relative error values. This figure clearly demonstrates the advantage of the different optimization models, $L^2$-$L^1$, $L^2$-$L^2$, $L^2$-$L^2L^1$, and $L^2$-$H^2$, over naive bilinear interpolation. Moreover, bilinear interpolation shows obvious disadvantages compared to all the results from the different models.

To enable a more detailed evaluation of the model performance, particularly avoiding the loss of detail caused by the wide dynamic range of the vertical axis, we focus primarily on the optimization models. If the bilinear interpolation results are included in the figure, it indicates that they are comparable to the outcomes of our optimization model. It is worth emphasizing that even if overlap occurs, the different marker shapes can be used to identify whether the bilinear information is included. If they are not drawn, it means that they are beyond the range of the drawing, indicating a significant deviation, which is considered poor performance in our context. Therefore, we do not pay more attention to them in the numerical results. Subsequent results all follow this rule.

\begin{figure}[htbp]
    \center
    \includegraphics[width=0.96\textwidth]{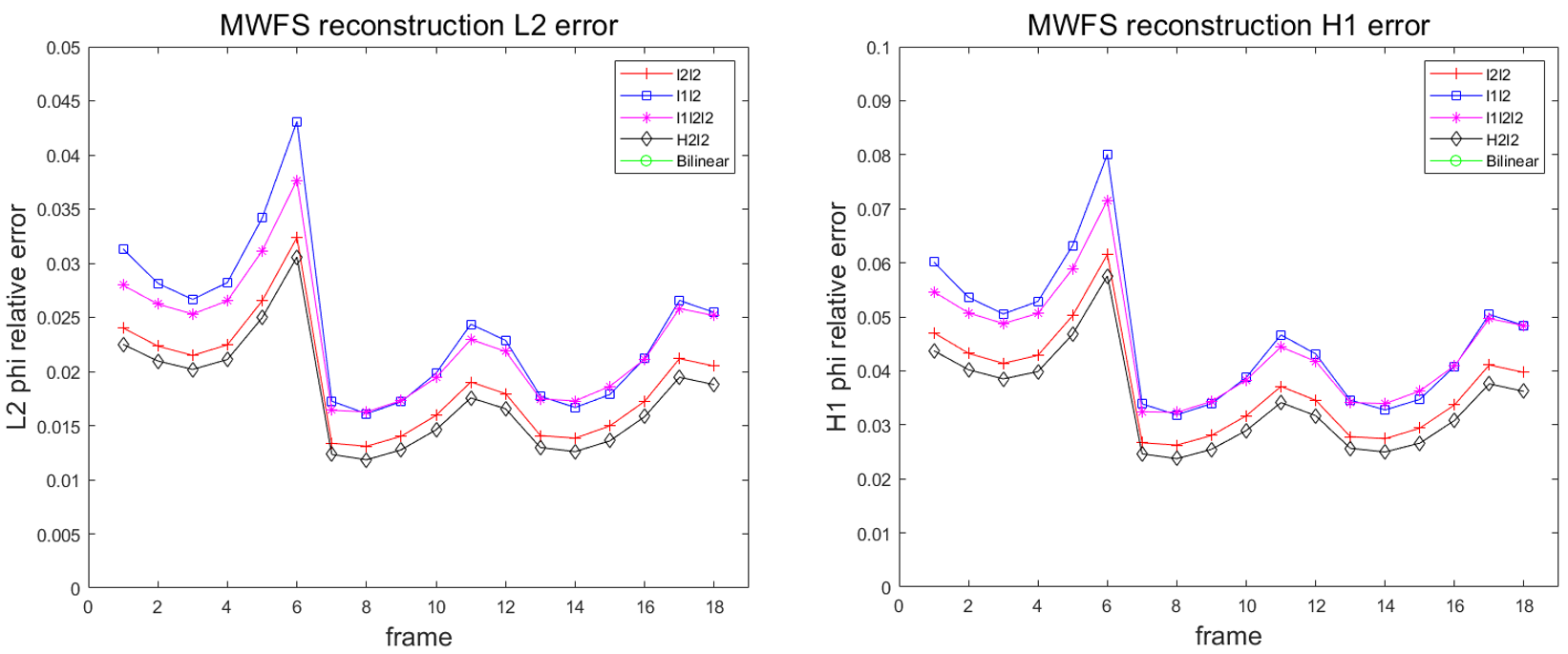}
    \caption{The $L_2$ relative error and $H_1$ relative error of the reconstruction. This is a zoom into Figure~\ref{fig: Bilinear_single}.}
    \label{fig: rr_single}
\end{figure}

In this simulation, Figure~\ref{fig: rr_single} provides detailed insights. The existing methods give reasonable results, but we still get an improvement. Generally, our method outperforms all other models, achieving the best reconstruction results in every frame. Specifically, our method yields the lowest relative errors with respect to both the $L_2$ and $H_1$ norms. 

In Table \ref{tab: s1gain}, we present the mean results for each model. Each column title denotes the name of a model. Compared to other models, our results demonstrate a significant enhancement, with an improvement ranging from $8.7\%$ to $24.6\%$ in relative error. Additionally, the ${H_1}$ relative error has shown an improvement of $8.5\%$ to $23.4\%$. These advancements underscore the effectiveness of our approach in optimizing the reconstruction process.

\begin{table}[htbp]
    \centering
    \caption{Mean Relative error}
    \begin{tabular}{|c|c|c|c|c|}
    \hline
       & $L^2$-$L^1$ & $L^2$-$L^2$ & $L^2$-$L^2L^1$ & $L^2$-$H^2$\\
    \hline
    Mean $err_{L_2}$ & 0.0242 & 0.0191 & 0.0230 & \textbf{0.0177}\\
    \hline
    Mean $err_{H_1}$ & 0.0461 & 0.0372 & 0.0446 & \textbf{0.0312}\\
    \hline
    \end{tabular}
    \label{tab: s1gain}
\end{table}

\begin{figure}[htbp]
    \center
    \includegraphics[width=0.62\textwidth]{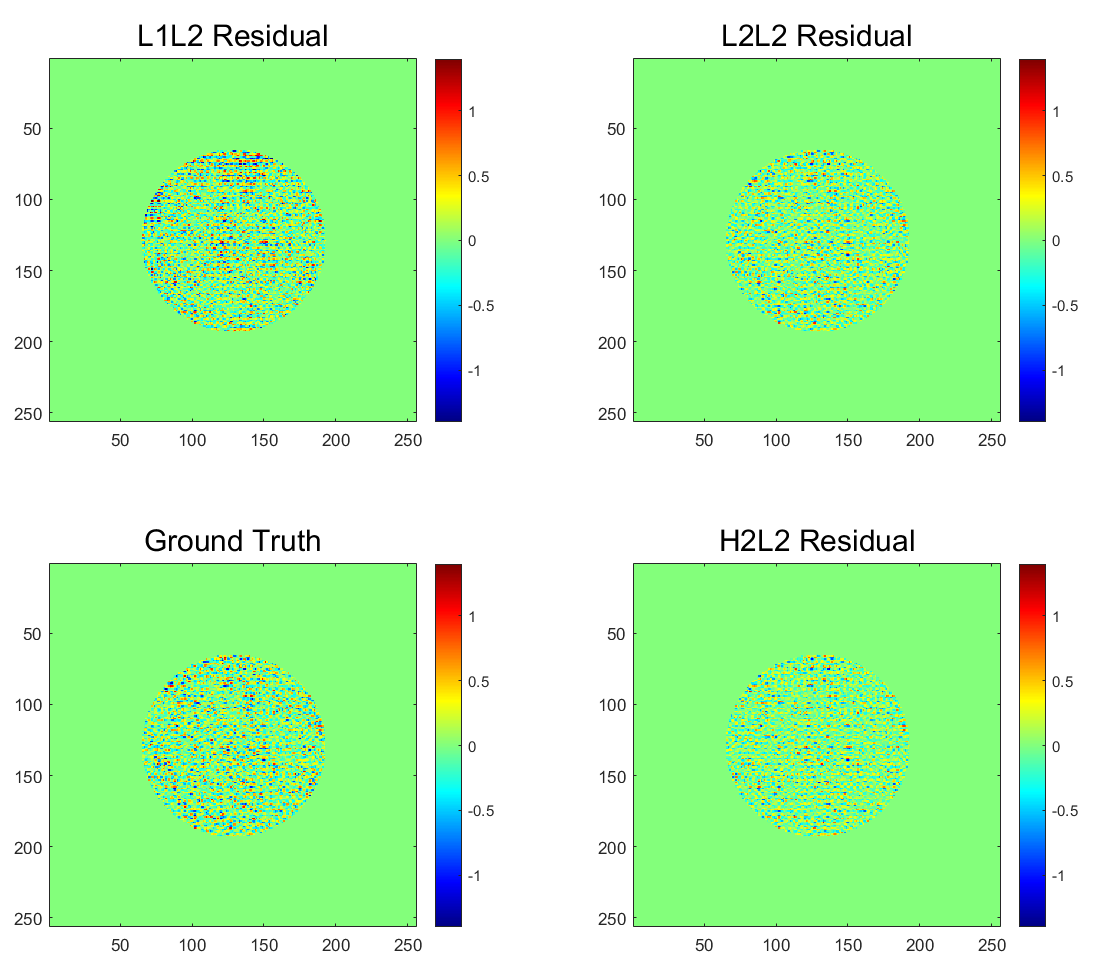}
    \caption{Visualization of phase residual errors for frame $3$.}
    \label{fig: Residual}
\end{figure}

In addition, Figure~\ref{fig: Residual} displays the phase residual errors. It is evident that the $H_2$ regularization term leads to a residual image in our model that is considerably lighter than others, indicating a superior reconstruction quality.

\begin{figure}[htbp]
    \center
    \includegraphics[width=0.90\textwidth]{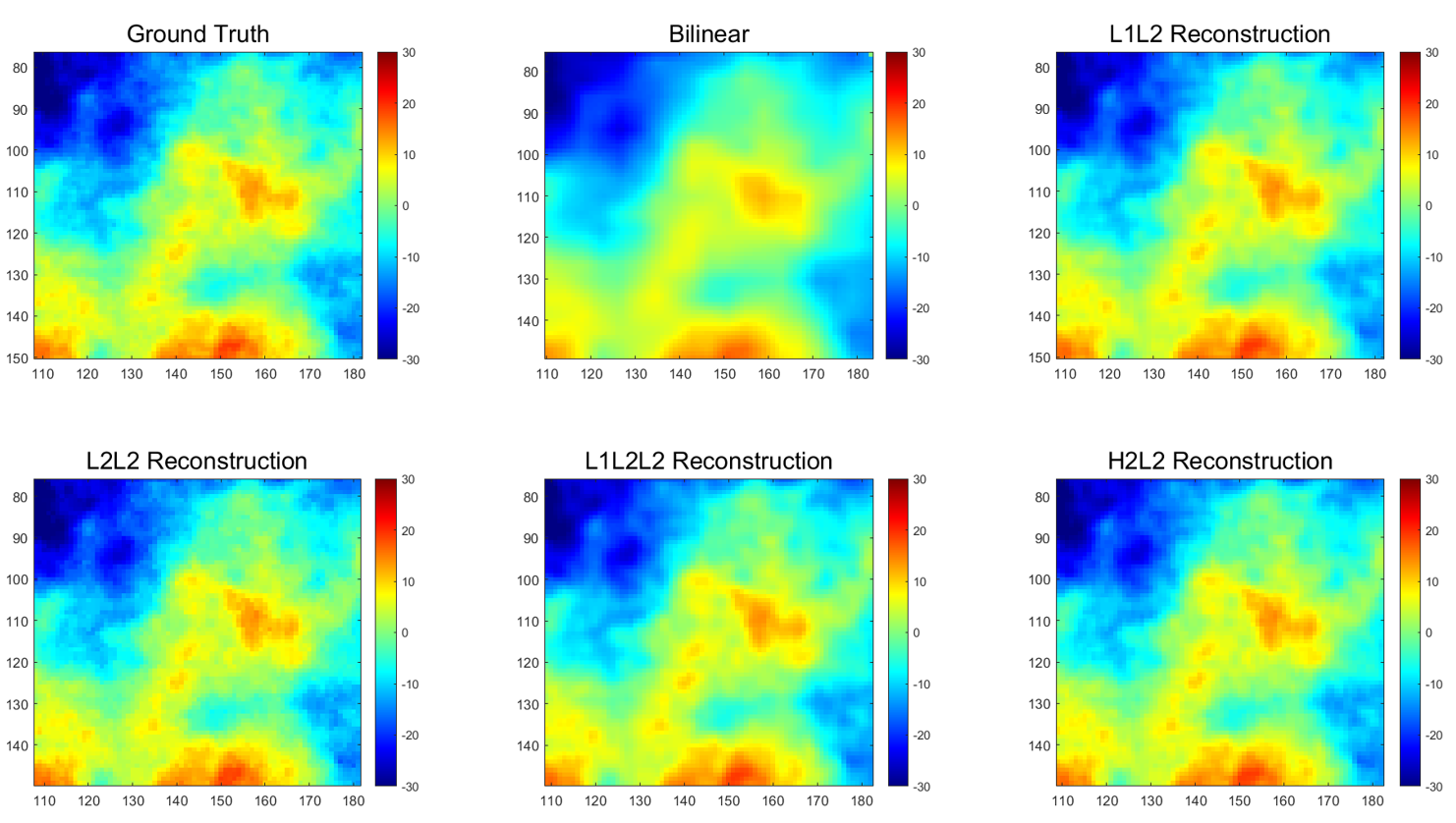}
    \caption{Magnified visualization of the reconstructed wavefront.}
    \label{fig: Phase_vis}
\end{figure}

To provide a clear visual comparison, Figure \ref{fig: Phase_vis} presents images of the reconstructed wavefront gradient using bilinear interpolation, the high-resolution ground-truth phase, and high-resolution reconstructions from both our proposed methods and those models we compare against. The bilinear interpolation method fails to accurately capture the wavefront phase information, and the performance of the $L^2$-$L^1$ model is also substantially worse compared to other models.

\begin{figure}[htbp]
    \center
    \includegraphics[width=0.88\textwidth]{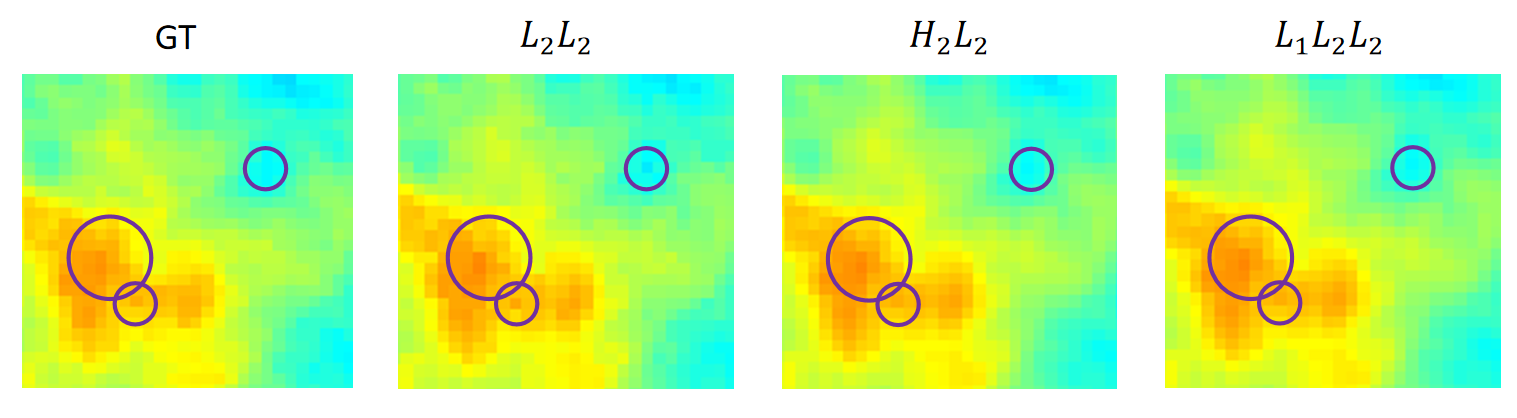}
    \caption{Detailed visualization of the reconstructed phase.}
    \label{fig: Detail_vis_full}
\end{figure}

In the analysis of other methods, we focus on the zoomed-in central region and present an enlarged view in Figure~\ref{fig: Detail_vis_full} to better illustrate fine-scale differences. It can be observed that the $L^2$-$L^2$ model introduces artificial features not present in the ground truth, indicating relatively high discretization error. In contrast, our $L^2$-$H^2$ model effectively preserves fine structures while avoiding such artifacts, leading to the most faithful reconstruction overall. The $L^2$-$L^2L^1$ model incorrectly preserves details that do not exist in the ground truth, particularly in the region marked by the largest annotation.

In conclusion, by applying slightly higher regularization than the statistical properties, the discretization error can be mitigated, and the reconstruction can be refined to be closer to reality.

\subsection{Numerical Consistency}\label{c3Consistency}
In this section, we demonstrate the robustness of our model under diverse practical scenarios by examining two distinct cases: first, fluctuating wind conditions with fixed WFS positions; and second, varying WFS positions under constant wind conditions.

\begin{figure}[htbp]
    \centering
    \includegraphics[width=0.78\textwidth]{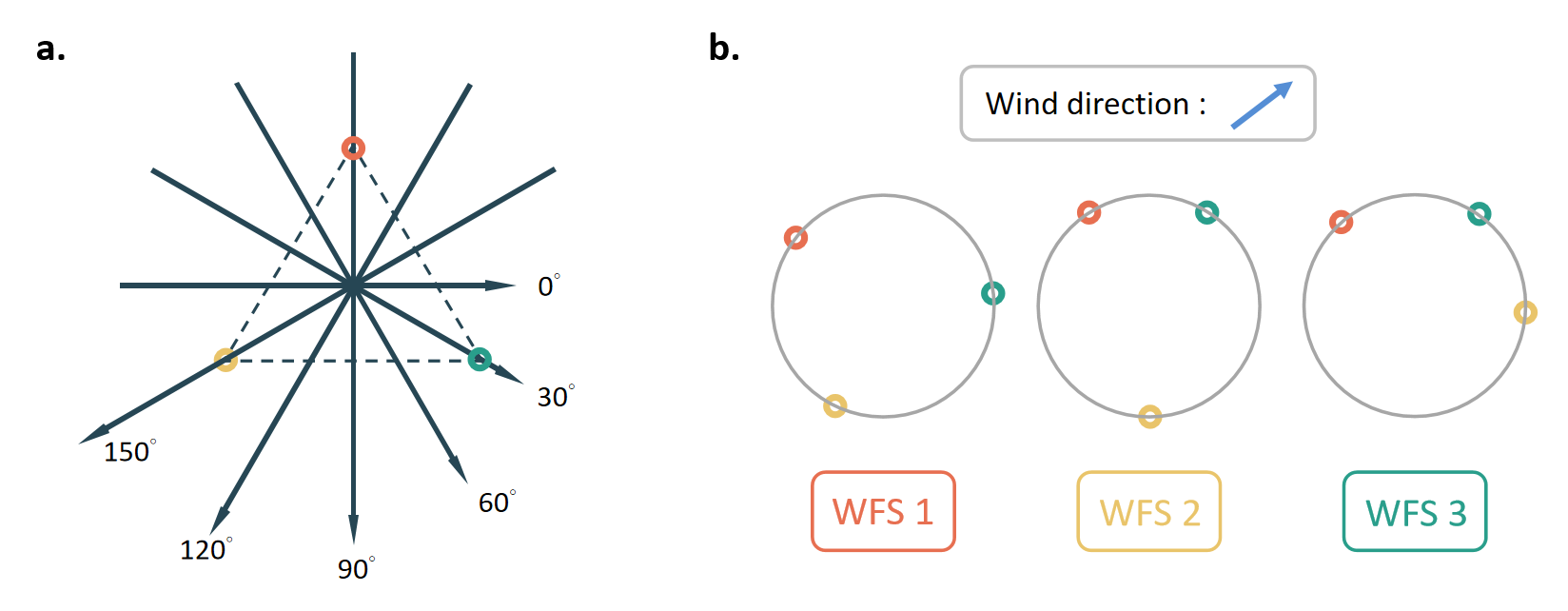}
    \caption{Two cases we considered in our estimation of our model. a) Examining how our model performed under different wind conditions, including variations in both the direction and magnitude of the wind. b) Testing our model's performance under various combinations of WFS.}
    \label{fig: Fixed}
\end{figure}

\subsubsection{Adaptability analysis to wind}
First, we evaluate our model's performance under varying wind direction and magnitude under the arrangement of the equilateral triangle, as shown in Figure~\ref{fig: Fixed}$\textbf{a}$, considering both the individual and average performance across fixed multi-WFS configurations. As for the implementation details, we select the representative directions at $0^\circ$, $30^\circ$, $60^\circ$, $90^\circ$, $120^\circ$, and $150^\circ$, paired with corresponding wind speed of $10 \ m/s$, $12.5 \ m/s$, $15 \ m/s$, $17.5 \ m/s$, $20 \ m/s$, $22.5 \ m/s$ respectively for illustration.

\begin{figure}[htbp]
    \centering
    \includegraphics[width=0.96\textwidth]{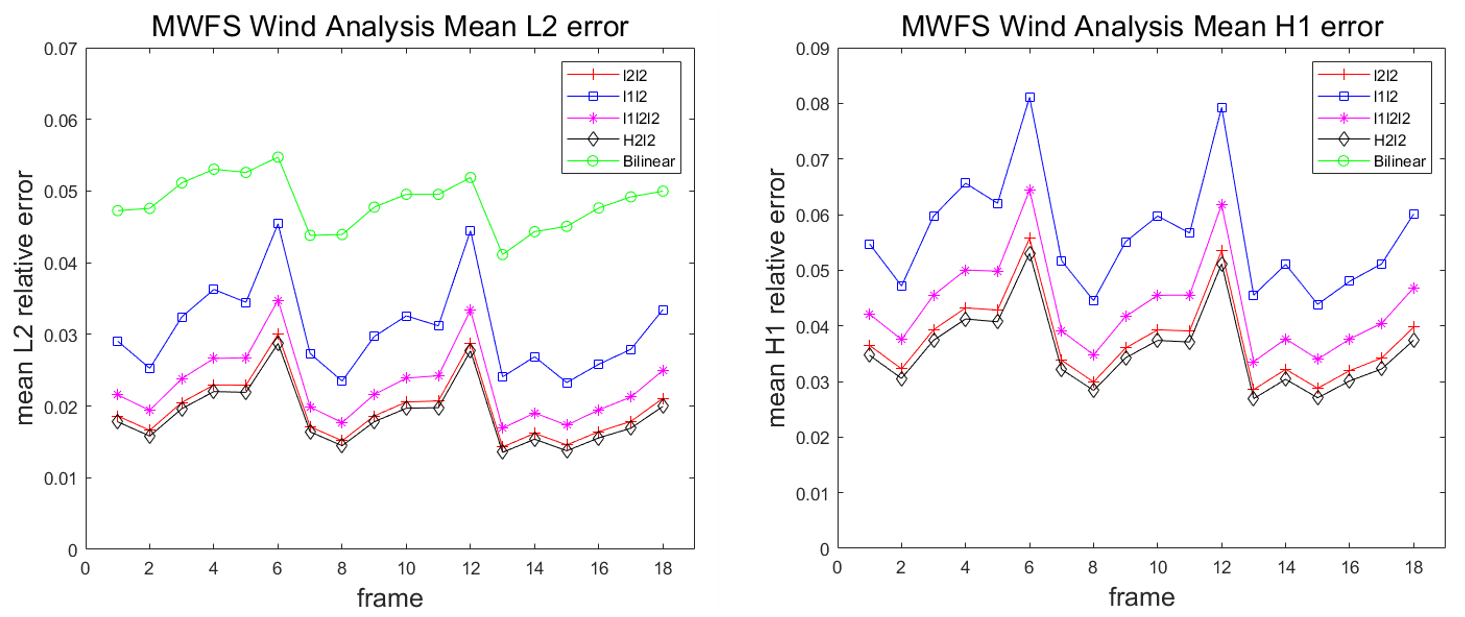}
    \caption{The mean $L_2$ and $H_1$ error under different wind directions and speeds.}
    \label{fig: rr_wind}
\end{figure}

The numerical results are depicted in Figure~\ref{fig: rr_wind}. The vertical axis indicates the mean relative error, calculated as the average of experimental outcomes across the six experimental setups, while the horizontal axis denotes the frame index. Notably, our approach exhibits the lowest error. Additionally, a discernible pattern is observed: all models, except bilinear interpolation, tend to exhibit higher errors at the beginning and end of the time steps cycle. This disparity arises from the fact that the frames in the middle of the sequence capture more information relative to those on the two sides. Previous Figure~\ref{fig: rr_single} also exhibits this pattern, but it is less pronounced.

\begin{figure}[htbp]
    \center
    \includegraphics[width=0.92\textwidth]{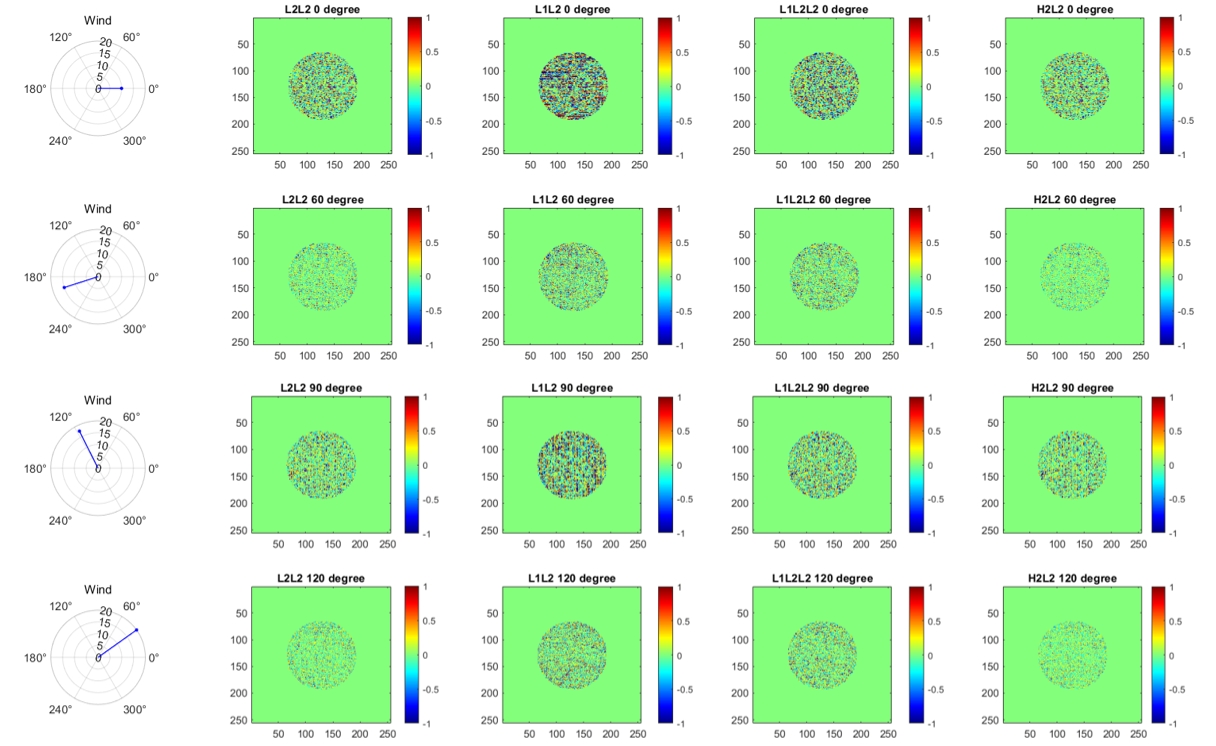}
    \caption{Visualization of phase residual errors for frame $3$ under wind directions at $0^{\circ}$, $60^{\circ}$, $90^{\circ}$ and $120^{\circ}$.}
    \label{fig: Residual_w_4}
\end{figure}

In Figure~\ref{fig: Residual_w_4}, we provide a comprehensive visualization of the phase residual errors resulting from reconstructions at wind angles of $0^\circ$, $60^\circ$, $90^\circ$, and $120^\circ$, using the $L^2$-$L^1$, $L^2$-$L^2$, $L^2$-$L^2L^1$, and our $L^2$-$H^2$ models.

Focus on the pattern of the residual visualization, especially at the $0^\circ$ and $90^\circ$ cases, a wind-aligned pattern is distinctly observable. Using $0^\circ$ as an example, the distinct horizontal lines indicate a relatively stable constant error along the horizontal axis. When examining our measurements of the wavefront gradients, the presence of a continuous horizontal constant error band indicates fewer deviations in horizontal gradient, but considerable discrepancies in vertical direction across the entire line. Essentially, models in horizontal wind conditions efficiently capture information along the horizontal axis, but significantly less along the vertical axis. This influence is particularly pronounced in the $L^2$-$L^1$ and $L^2$-$L^2L^1$ models due to the $L_1$ regularization term. This wind pattern is less distinct at $60^\circ$ and $120^\circ$ degrees since the provision of adequate information along both axes. Our method also demonstrates superior performance in these configurations compared to the $0^\circ$ and $90^\circ$ configurations.

In general, by assessing the model's performance across all mentioned $4$ experimental configurations, our model consistently exhibits the slightest residual error. We gain insight into the robustness of our $L^2$-$H^2$ model under diverse wind conditions.

\subsubsection{WFS position analysis}
Furthermore, we assess the performance of the proposed models under uniform wind conditions across different multi-WFS combinations as illustrated in Figure~\ref{fig: Fixed}$\textbf{b}$. To ensure the credibility of the results, we evaluate them under two different wind scenarios, where the wind direction for one is parallel to the horizontal, and the other is at an angle of $30^\circ$ to the horizontal.

\begin{figure}[htbp]
    \center
    \includegraphics[width=0.88\textwidth]{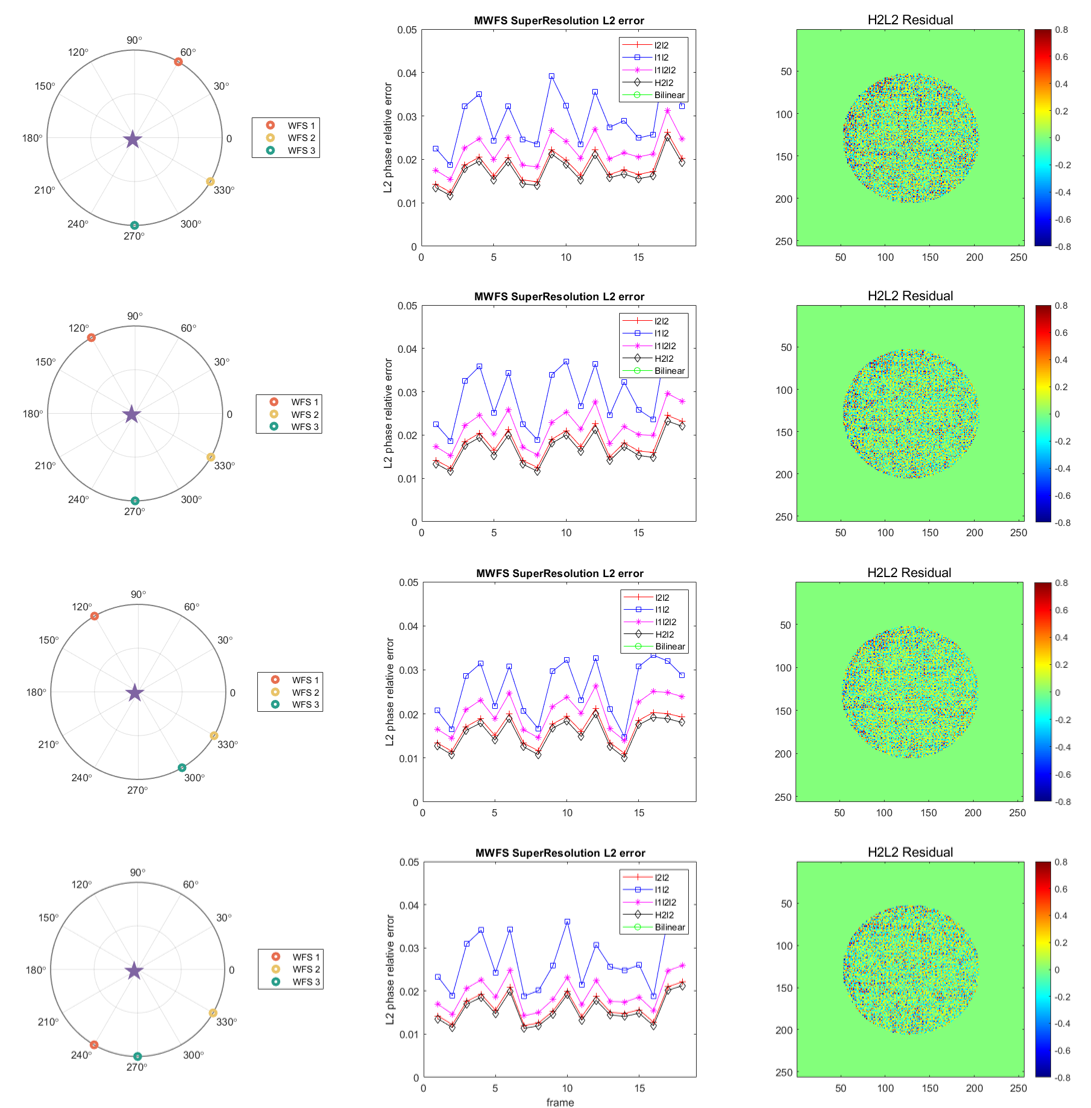}
    \caption{The reconstruction results under various WFS positions are shown for a wind speed of $15 \ m/s$ in the $0^{\circ}$ direction. The left column depicts the visual arrangement of WFS positions, the middle column exhibits the relative error for each frame among all models, and the right column offers a visualization of the phase residual error for the $3$-th frame, specific to our model.}
    \label{fig: Pos_analysis_1}
\end{figure}

\begin{figure}[htbp]
    \center
    \includegraphics[width=0.88\textwidth]{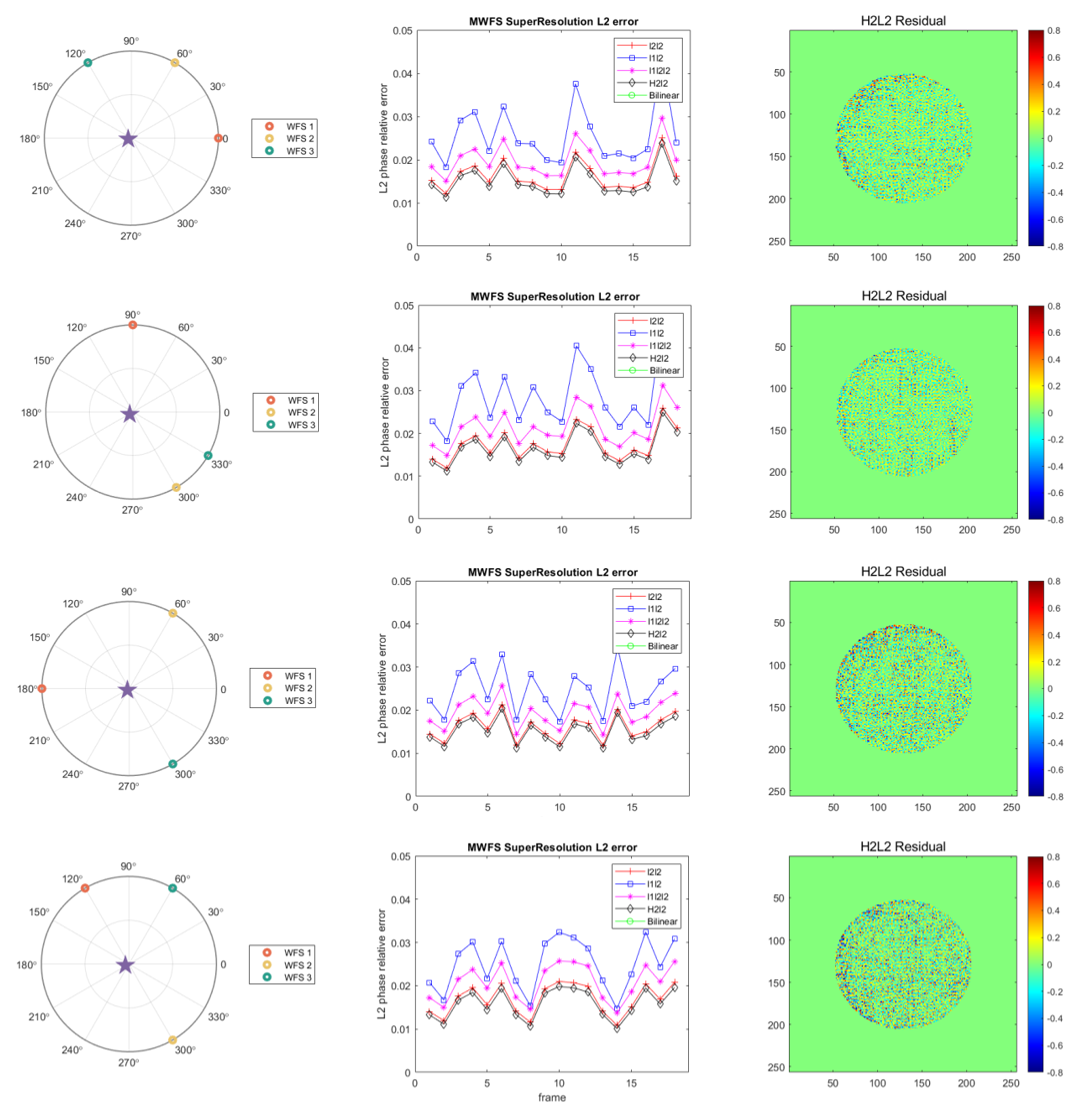}
    \caption{The reconstruction results under various WFS positions are shown for a wind speed of $15 \ m/s$ in the $30^{\circ}$ direction. The left column depicts the visual arrangement of WFS positions, the middle column exhibits the relative error for each frame among all models, and the right column offers a visualization of the phase residual error for the $3$-th frame, specific to our model.}
    \label{fig: Pos_analysis_2}
\end{figure}

In our study, we selected four distinctly different positions for guide stars and thus WFSs. Related numerical results are given in Figure~\ref{fig: Pos_analysis_1} and Figure~\ref{fig: Pos_analysis_2} for $0^\circ$  and $30^\circ$ wind direction, respectively. The numerical findings clearly demonstrate that, under the two wind configurations discussed and varying WFS positions, our $L^2$-$H^2$ model consistently outperforms the others. Additionally, the phase residual errors show no significant flaws. Since all models assign equal weight to each measurement, these results can be extended to other positions by symmetry considerations.

Overall, by assessing the model's performance across the above two experiments, the numerical and visual findings indicate that our model is robust and consistently achieves high accuracy in recovering information across various scenarios. Our $L^2$-$H^2$ model exhibits robustness and showcases the superior ability in atmospheric reconstruction.

\subsection{Reference frame analysis} \label{c3Refframe}
\begin{figure}[htbp]
    \centering
    \includegraphics[width=0.84\textwidth]{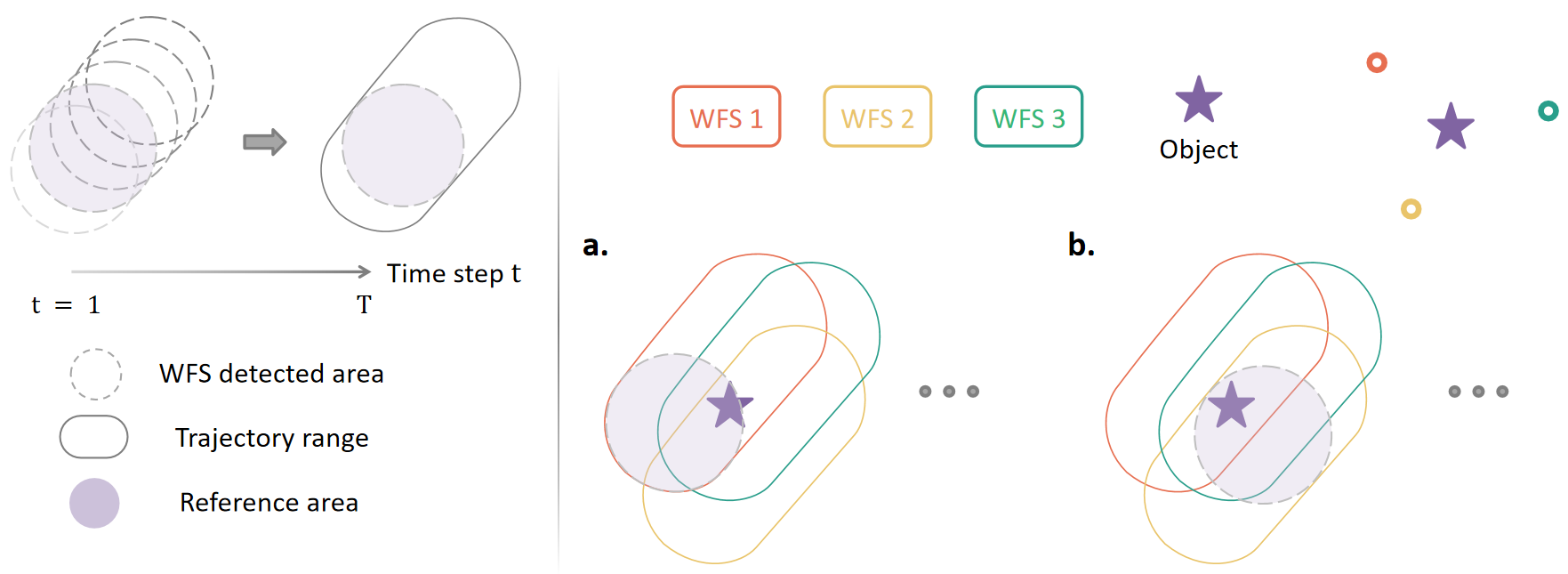}
    \caption{On the left hand, explains the visualization of the reference frame and the trajectory range. On the right side, the position of the WFS and a series of options of reference frames are depicted, $\textbf{a}$: Utilizing frame $1$ as the reference frame. $\textbf{b}$: Utilizing frame $10$ as the reference. }
    \label{fig: Ref_Frame_Frame_title}
\end{figure}

As described in \eqref{WDRA}, we can model the multiple WFSs under different reference frames. In this section, we analyze the reconstruction effect of the fidelity term established regarding the different frames under the same observation as illustrated in Figure~\ref{fig: Ref_Frame_Frame_title}. To highlight the distinctions between each frame, we employ a dataset that incorporates more high-frequency information.

\begin{figure}[htbp]
    \centering
    \includegraphics[width=0.84\textwidth]{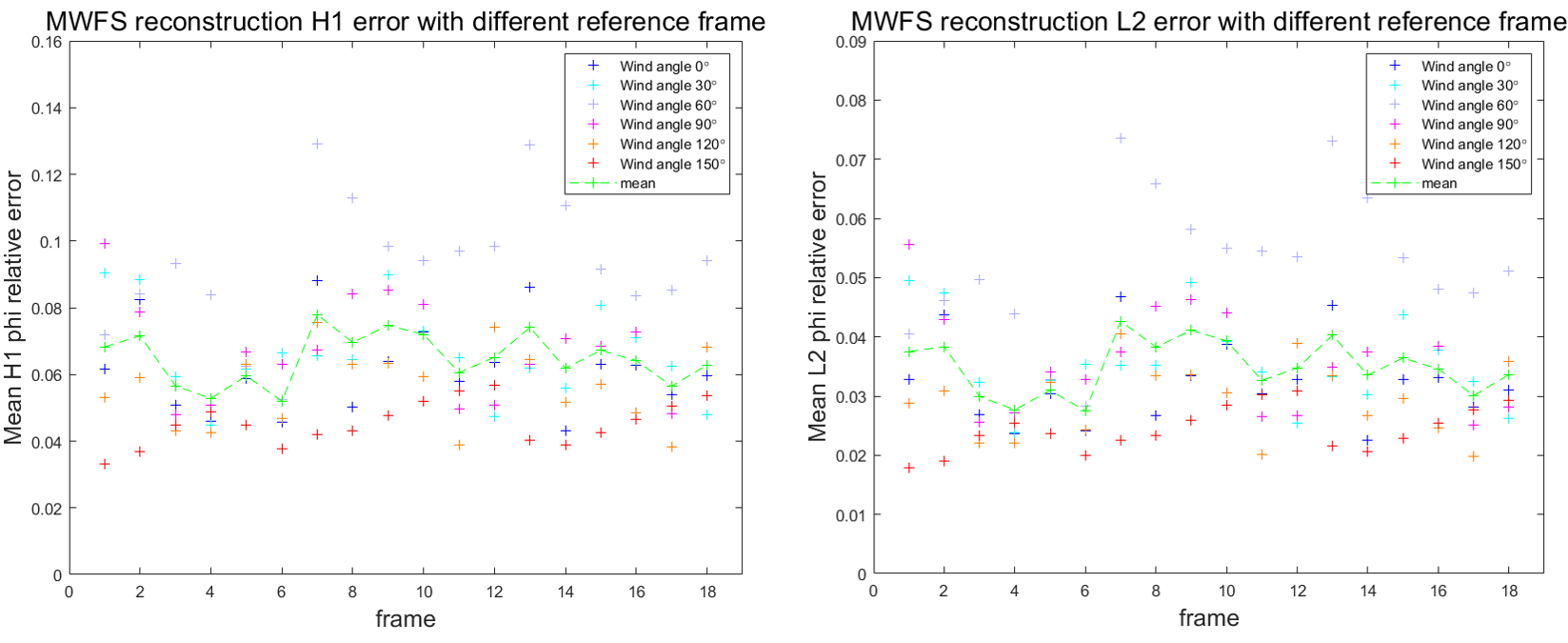}
    \caption{The plots show the mean relative error among all frames under different reference frames. Different colors represent the mean performance among all frames under the corresponding $x$-axis frame index as reference. The light green line marked as ``mean'' represents the average performance among all wind directions.}
    \label{fig: rr_ref_Frames_1}
\end{figure}

Figure~\ref{fig: rr_ref_Frames_1} shows the average reconstruction results over $18$ frames at a wind speed of $15 \ m/s$ in different directions and the mean performance over all the directions. The horizontal axis denotes the reference frame index, i.e., results on one point indicate a result based on the horizontal axis index frame as the reference in our model, and the vertical axis represents the relative error. The six different colors correspond to the average error of all frames under six different wind directions. The light green line marked as ``mean'' is the average of the six experiments, i.e., the average of the average error of all frames under the six wind directions obtained by taking the frame corresponding to the horizontal axis index as the reference. In general, the selection of frames influences the final reconstruction outcome, though the impact is relatively minor.

\subsection{Single and multiple WFS strategy comparison}\label{c3singlemulti}
\begin{figure}[htbp]
    \center
    \includegraphics[width=0.82\textwidth]{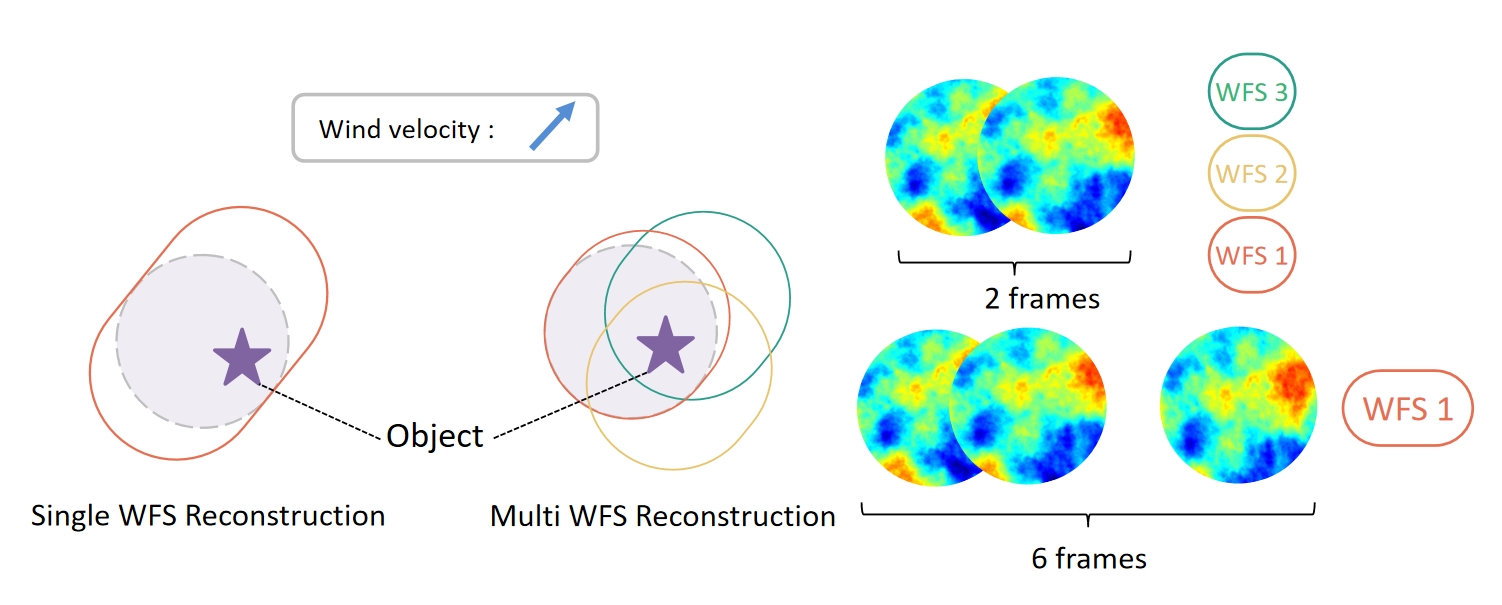}
    \caption{The comparison between single and multi-WFS strategies is demonstrated using a scenario involving $6$ frames in total. In this context, the trajectory that closely resembles a perfect circle corresponds to the detected area in multi-WFS experiments, while the more elongated trajectory represents the detected area using a single WFS with a triple number of frames.}
    \label{fig: Ref_Single_Multi_WFS}
\end{figure}

This section presents a comparative analysis between single and multi-WFS systems in our $L^2$-$H^2$ model, specifically designed to validate the benefits of employing multi-WFS strategies. To guarantee a fair comparison and maintain alignment, simulation parameters for both the multi-WFS and single-WFS scenarios were kept consistent, except that the time step for the single-WFS case was tripled to achieve an equivalent total number of detected frames. Subsequently, to ensure that the detection frames of a single WFS cover the target object, wind with a speed of $8~m/s$ is utilized.  The experimental setup is illustrated in Figure~\ref{fig: Ref_Single_Multi_WFS}.

We judge the reconstruction performance of single-WFS and multi-WFS under different noise levels and wind settings. The proposed method was evaluated in $18$ frames under $18$ time steps for a single WFS, and $6$ time steps for $3$ WFSs. 

\begin{figure}[htbp]
    \center
    \includegraphics[width=0.78\textwidth]{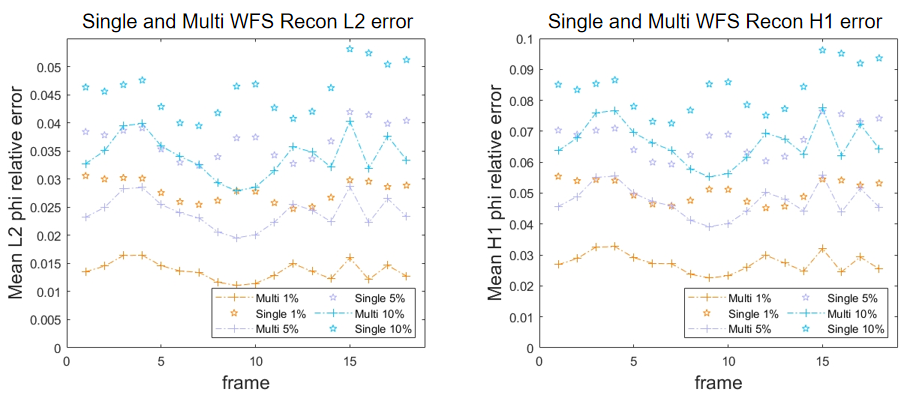}
    \caption{Mean $L_2$ relative error and $H_1$ relative error under different wind conditions in different noise levels.}
    \label{fig: Single_multi_noise_0.1}
\end{figure}

First, we conducted a comparative analysis of the system's performance under $1\%$, $5\%$, $10\%$ noise levels, with the mean reconstruction error under $6$ wind directions depicted in Figure \ref{fig: Single_multi_noise_0.1}. In this figure, the discrete markers represent the performance of the single-WFS, while the lines illustrate the performance of the multi-WFS system. Across every color representing various noise levels, the multi-WFS consistently achieves better results than the single-WFS. This holds even under challenging conditions, such as with a significant noise level of $10\%$. As expected, the multi-WFS system exhibits significantly enhanced performance over the single-WFS system at an appropriate noise level, confirming the robustness of our approach in different noise levels. 

\begin{figure}[htbp]
    \center
    \includegraphics[width=0.92\textwidth]{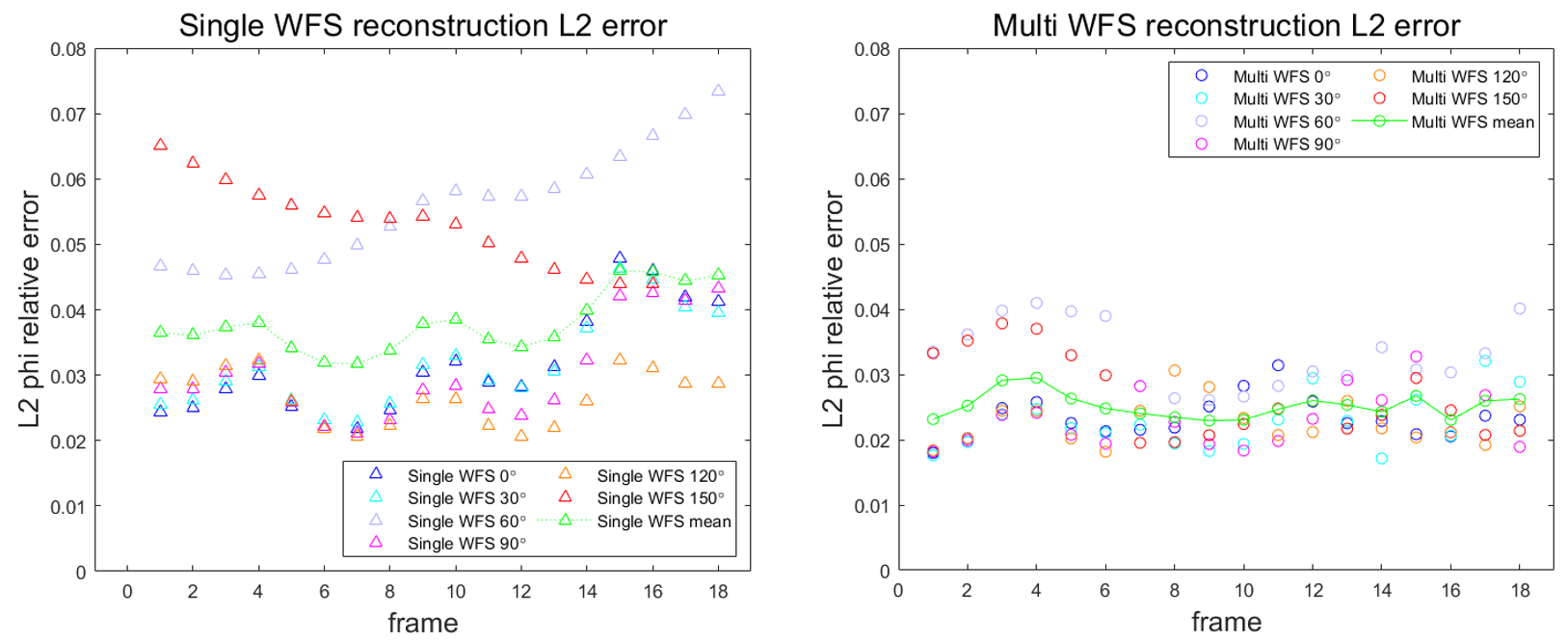}
    \caption{$L_2$ relative error among all the detected areas under various wind directions. The left figure represents the single-WFS mode, and the right figure indicates the multi-WFS mode. Both figures have the same vertical scale to emphasize the difference in model performance between the two modes.}
    \label{fig: Single_Multi_wind_2}
\end{figure}

Then, we compare the performance of single-WFS and multi-WFS systems under different wind directions with $5\%$ noise level, as illustrated in Figure~\ref{fig: Single_Multi_wind_2}. 

The figures on the left and right show the $L_2$ relative errors across a range of wind directions for single-WFS and multi-WFS configurations, respectively. Discrete markers in different colors, specifically triangles for single-WFS and circles for multi-WFS, represent their respective performances. The line graph depicts the average error across all wind directions. It is worth noting that the vertical axes of both graphs are scaled identically to facilitate a clear comparison between the single and multi-WFS configurations. 

By observing the overall data distribution in both the left and right figures in Figure~\ref{fig: Single_Multi_wind_2}, it is evident that the results for the multi-WFS (right figures) are consistently lower than those for the single-WFS (left figures). This indicates that the multi-WFS system has a significant stability advantage under different wind conditions.

Overall, our findings consistently highlight the superior performance of the multi-WFS system over its single-WFS counterpart, across all tested conditions, including various upsampling factors, wind directions, and magnitudes. The enhanced positional relationships among the WFSs significantly contribute to the increased stability and precision in wavefront reconstruction.

\subsection{Computational cost}
In this section, we evaluated the runtime performance of our approach. Our method demonstrates fast execution speeds. Table~\ref{Tab: CPU} shows the CPU time in seconds (s). This clearly shows that our method also outperforms the others in run time.

\begin{table}[htbp]
    \centering
    \caption{CPU time costing}
    \begin{tabular}{|c|c|c|c|c|}
    \hline
      & $L^2$-$L^1$ & $L^2$-$L^2$ & $L^2$-$L^2L^1$ & $L^2$-$H^2$\\
    \hline
    CPU time (s) & 2.9097 & 27.2853 & 32.5108 & \textbf{2.8404} \\
    \hline
    \end{tabular}
    \label{Tab: CPU}
\end{table}

To illustrate the computational cost and memory usage, we report them together with computational details for reconstructions on a $256 \times 256$ grid with 3 WFS and 6 time steps under different upsampling factors. For upsampling factor 2, the system matrix contains $2.10 \times 10^6$ non-zero entries (fill ratio $0.0488 \%$) and requires 36.44 MB of memory; each PCG iteration involves $4.19 \times 10^6$ FLOPs, converging in 1000 iterations within 2.18 s. For upsampling factor 4, the matrix has $5.20 \times 10^6$ non-zero entries (fill ratio $0.121 \%$), 84.02 MB memory, $1.04 \times 10^7$ FLOPs per iteration, and converges in 991 iterations within 4.84 s. 

Given the problem size in our study, the system matrices are highly sparse, symmetric, and positive-definite, which allows the conjugate gradient iterations to converge efficiently. Consequently, the reconstruction can be completed within a few seconds on a standard workstation. Although our focus is on high-quality post-processed results based on multiple frames rather than real-time execution, the sparsity of the system matrices ensures that the computation remains fast enough to be practical within the context of our study.

\section{Conclusion}
We introduce a new model using statistical properties of atmospheric turbulence to enhance the computation of a super-resolution wavefront in ground-based astronomy. To reduce the computational cost, we approximate $H^{11/6}$ by the smoother space $H^2$, which also ensures that discretization noise is damped. We compare our model to existing methods for super-resolution wavefront reconstruction.

In summary, our $L^2$-$H^2$ model effectively captures finer details of the atmospheric wavefront and demonstrates robust stability across various experimental conditions, making it well-suited for practical deployment in dynamic atmospheric environments. Furthermore, the multi-WFS reconstruction not only accelerates detection efficiency but also exploits the inherent spatial relationships among WFSs to enhance reconstruction accuracy, outperforming single-WFS reconstruction when using an equivalent number of frames.

In this paper, we consider only single-layer atmospheric turbulence. As future work, we will extend our model to multi-layer atmospheric turbulence and conduct stress tests under multiple layers and inexact wind speeds. Moreover, tailored preconditioners for the conjugate gradient solver could be developed to accelerate convergence in large-scale problems. Additionally, we plan to test our method using end-to-end simulation tools for AO systems to further verify its performance. Subsequently, we aim to use the super-resolution wavefront as input for point spread function (PSF) reconstruction methods, such as those presented in \cite{WaHoRa18, WaNiRa23, Wagner_2022}.

\section*{Funding}
The work of YW, RW, and RR was supported by the Austrian Science Fund (FWF) under project F6805-N36, SFB Tomography Across the Scales. The work of RC was supported by the Hong Kong Research Grants Council (HKRGC), Grant No. CityU11309922, the Innovation and Technology Fund (ITF), Grant No. MHP/054/22, and the LU BGR Grant No. 105824.

\printbibliography

@article{Shack71,
  author = {Ronald Shack},
  year = {1971},
  title = {Production and use of a lenticular {H}artmann screen},
  journal = {J. Opt. Soc. Am.},
  volume = {61},
  number = {656},
  pages = { }
}

@article{WaHoRa18,
author = {Roland Wagner and Christoph Hofer and Ronny Ramlau},
title = {Point spread function reconstruction for Single-conjugate Adaptive Optics}, 
journal = {Journal of Astronomical Telescopes, Instruments, and Systems},
year = {2018},
volume = {4},
number = {4},
pages = {049003},
doi = {10.1117/1.JATIS.4.4.049003}
}

@article{WaNiRa23,
author = {Roland Wagner and Jenny Niebsch and Ronny Ramlau},
journal = {J. Opt. Soc. Am. A},
number = {7},
pages = {1382--1391},
publisher = {Optica Publishing Group},
title = {Off-axis point spread function reconstruction for single conjugate adaptive optics},
volume = {40},
month = {Jul},
year = {2023},
url = {https://opg.optica.org/josaa/abstract.cfm?URI=josaa-40-7-1382},
doi = {10.1364/JOSAA.488843}
}

@book{roggemann2018imaging,
  title={Imaging through turbulence},
  author={Roggemann, Michael C and Welsh, Byron M},
  year={2018},
  publisher={CRC press}
}

@book{adams2003sobolev,
  title={Sobolev spaces},
  author={Adams, Robert A and Fournier, John JF},
  year={2003},
  publisher={Elsevier}
}

@article{di2012hitchhikerʼs,
  title={Hitchhiker's guide to the fractional Sobolev spaces},
  author={Di Nezza, Eleonora and Palatucci, Giampiero and Valdinoci, Enrico},
  journal={Bulletin des sciences math{\'e}matiques},
  volume={136},
  number={5},
  pages={521--573},
  year={2012},
  publisher={Elsevier}
}

@inproceedings{bharmal2015frozen,
  title={Frozen flow or not? Investigating the predictability of the atmosphere},
  author={Bharmal, Nazim Ali},
  booktitle={Journal of Physics: Conference Series},
  volume={595},
  number={1},
  pages={012003},
  year={2015},
  organization={IOP Publishing}
}

@article{bardsley2008wavefront,
  title={Wavefront reconstruction methods for adaptive optics systems on ground-based telescopes},
  author={Bardsley, Johnathan M},
  journal={SIAM Journal on Matrix Analysis and Applications},
  volume={30},
  number={1},
  pages={67--83},
  year={2008},
  publisher={SIAM}
}

@book{goodman2005introduction,
  title={Introduction to Fourier optics},
  author={Goodman, Joseph W},
  year={2005},
  publisher={Roberts and Company publishers}
}

@misc{platt2001history,
  title={History and principles of Shack-Hartmann wavefront sensing},
  author={Platt, Ben C and Shack, Ronald},
  journal={Journal of refractive surgery},
  volume={17},
  number={5},
  pages={S573--S577},
  year={2001},
  publisher={SLACK Incorporated Thorofare, NJ}
}

@book{davidson2015turbulence,
  title={Turbulence: an introduction for scientists and engineers},
  author={Davidson, Peter Alan},
  year={2015},
  publisher={Oxford university press}
}

@incollection{roddier1981v,
  title={The effects of atmospheric turbulence in optical astronomy},
  author={Roddier, Fran{\c{c}}ois},
  booktitle={Progress in optics},
  volume={19},
  pages={281--376},
  year={1981},
  publisher={Elsevier}
}

@article{ke2020reconstruction,
  title={Reconstruction of the high resolution phase in a closed loop adaptive optics system},
  author={Ke, Rihuan and Wagner, Roland and Ramlau, Ronny and Chan, Raymond},
  journal={SIAM Journal on Imaging Sciences},
  volume={13},
  number={2},
  pages={775--806},
  year={2020},
  publisher={SIAM}
}

@article{chan2013phase,
  title={A phase model for point spread function estimation in ground-based astronomy},
  author={Chan, Raymond Honfu and Yuan, XiaoMing and Zhang, WenXing},
  journal={Science China Mathematics},
  volume={56},
  pages={2701--2710},
  year={2013},
  publisher={Springer}
}

@article{zou2005regularization,
  title={Regularization and variable selection via the elastic net},
  author={Zou, Hui and Hastie, Trevor},
  journal={Journal of the Royal Statistical Society Series B: Statistical Methodology},
  volume={67},
  number={2},
  pages={301--320},
  year={2005},
  publisher={Oxford University Press}
}

@inproceedings{nagy2010fast,
  title={Fast PSF reconstruction using the frozen flow hypothesis},
  author={Nagy, James and Jefferies, Stuart and Chu, Qing},
  booktitle={Proceedings of the Advanced Maui Optical and Space Surveillance Technologies Conference},
  pages={13--16},
  year={2010}
}

@article{taylor1938spectrum,
  title={The spectrum of turbulence},
  author={Taylor, Geoffrey Ingram},
  journal={Proceedings of the Royal Society of London. Series A-Mathematical and Physical Sciences},
  volume={164},
  number={919},
  pages={476--490},
  year={1938},
  publisher={The Royal Society}
}

@Article{Helin_2013,
  author    = {Tapio Helin and Mykhaylo Yudytskiy},
  journal   = {Inverse Problems},
  title     = {Wavelet methods in multi-conjugate adaptive optics},
  year      = {2013},
  month     = {jul},
  number    = {8},
  pages     = {085003},
  volume    = {29},
  doi       = {10.1088/0266-5611/29/8/085003},
  publisher = {{IOP} Publishing},
}

@article{wyngaard1992atmospheric,
  title={Atmospheric turbulence},
  author={Wyngaard, John C},
  journal={Annual Review of Fluid Mechanics},
  volume={24},
  number={1},
  pages={205--234},
  year={1992},
  publisher={Annual Reviews 4139 El Camino Way, PO Box 10139, Palo Alto, CA 94303-0139, USA}
}

@article{davies2012adaptive,
  title={Adaptive optics for astronomy},
  author={Davies, Richard and Kasper, Markus},
  journal={Annual Review of Astronomy and Astrophysics},
  volume={50},
  number={1},
  pages={305--351},
  year={2012},
  publisher={Annual Reviews}
}

@book{roddier1999adaptive,
  title={Adaptive optics in astronomy},
  author={Roddier, Fran{\c{c}}ois},
  year={1999},
  publisher={Cambridge university press}
}

@article{goodman1985statistical,
  title={Statistical optics},
  author={Goodman, Joseph W},
  journal={Journal of the Optical Society of America A},
  volume={2},
  number={9},
  pages={1448--1454},
  year={1985}
}

@book{tennekes1972first,
  title={A first course in turbulence},
  author={Tennekes, Hendrik and Lumley, John Leask},
  year={1972},
  publisher={MIT press}
}

@article{tatarskii1971effects,
  title={The effects of the turbulent atmosphere on wave propagation},
  author={Tatarskii, Valerian Ilitch},
  journal={Jerusalem: Israel Program for Scientific Translations, 1971},
  year={1971}
}

@book{rudin1964principles,
  title={Principles of mathematical analysis},
  author={Rudin, Walter and others},
  volume={3},
  year={1964},
  publisher={McGraw-hill New York}
}

@article{cai2013two,
  title={A two-stage image segmentation method using a convex variant of the Mumford--Shah model and thresholding},
  author={Cai, Xiaohao and Chan, Raymond and Zeng, Tieyong},
  journal={SIAM Journal on Imaging Sciences},
  volume={6},
  number={1},
  pages={368--390},
  year={2013},
  publisher={SIAM}
}

@article{hill1978spectra,
  title={Spectra of fluctuations in refractivity, temperature, humidity, and the temperature-humidity cospectrum in the inertial and dissipation ranges},
  author={Hill, Reginald J},
  journal={Radio Science},
  volume={13},
  number={6},
  pages={953--961},
  year={1978},
  publisher={Wiley Online Library}
}

@article{rousset1999wave,
  title={Wave-front sensors},
  author={Rousset, Gerard},
  journal={Adaptive optics in astronomy},
  volume={1},
  pages={91},
  year={1999}
}

@article{bose1998high,
  title={High-resolution image reconstruction with multisensors},
  author={Bose, NK and Boo, KJ},
  journal={International Journal of Imaging Systems and Technology},
  volume={9},
  number={4},
  pages={294--304},
  year={1998},
  publisher={Wiley Online Library}
}

@article{chan2003wavelet,
  title={Wavelet algorithms for high-resolution image reconstruction},
  author={Chan, Raymond H and Chan, Tony F and Shen, Lixin and Shen, Zuowei},
  journal={SIAM Journal on Scientific Computing},
  volume={24},
  number={4},
  pages={1408--1432},
  year={2003},
  publisher={SIAM}
}

@article{tsai1984multiframe,
  title={Multiframe image restoration and registration},
  author={Tsai, Roger Y and Huang, Thomas S},
  journal={Multiframe image restoration and registration},
  volume={1},
  pages={317--339},
  year={1984}
}

@article{racine1996telescope,
  title={The telescope point spread function},
  author={Racine, Rene},
  journal={Publications of the Astronomical Society of the Pacific},
  volume={108},
  number={726},
  pages={699},
  year={1996},
  publisher={IOP Publishing}
}

@article{hill1992review,
  title={Review of optical scintillation methods of measuring the refractive-index spectrum, inner scale and surface fluxes},
  author={Hill, Reginald J},
  journal={Waves in Random Media},
  volume={2},
  number={3},
  pages={179},
  year={1992},
  publisher={IOP Publishing}
}

@book{von1931mechanical,
  title={Mechanical similitude and turbulence},
  author={Von K{\'a}rm{\'a}n, Theodore},
  number={611},
  year={1931},
  publisher={National Advisory Committee for Aeronautics}
}

@book{chan2005image,
  title={Image processing and analysis: variational, PDE, wavelet, and stochastic methods},
  author={Chan, Tony F and Shen, Jianhong},
  year={2005},
  publisher={SIAM}
}

@inproceedings{tikhonov1943stability,
  title={On the stability of inverse problems},
  author={Tikhonov, Andrey Nikolayevich and others},
  booktitle={Dokl. akad. nauk sssr},
  volume={39},
  number={5},
  pages={195--198},
  year={1943}
}

@article{Fusco,
 author               = {Thierry Fusco and  Jean-Marc Conan and Gerard Rousset and Laurent Marc Mugnier and Vincent Michau},
 journal              = {J. Opt. Soc. Am. A},
 volume             = {18},
 number               = {10},
 pages                = {2527-2538},
 title                = {Optimal wave-front reconstruction strategies for multi conjugate adaptive optics},
 year                 = {2001},
 }

@article{YuHeRa13b,
 author               = {Mykhaylo Yudytskiy and Tapio Helin and Ronny Ramlau},
 journal 		= {J. Opt. Soc. Am. A},
 month 		= {Mar},
 number 		= {3},
 pages 		= {550--560},
 publisher 	= {OSA},
 title 			= {Finite element-wavelet hybrid algorithm for atmospheric tomography},
 volume 		= {31},
 year 		= {2014},
 url 			= {http://josaa.osa.org/abstract.cfm?URI=josaa-31-3-550},
 doi 			= {10.1364/JOSAA.31.000550}
}

@article{StadlerRamlau2021,
  author       = {Ramlau, Ronny and Stadler, Bernadett},
  title        = {An augmented wavelet reconstructor for atmospheric tomography},
  journal      = {Electronic Transactions on Numerical Analysis (ETNA)},
  pages        = {256--275},
  volume       = {54},
  year         = {2021},
}

@inproceedings{10.1117/12.2628969short,
author = {Paolo Ciliegi and Guido Agapito and Matteo Aliverti et al.},
title = {{MAORY/MORFEO at ELT: general overview up to the preliminary design and a look towards the final design}},
volume = {12185},
booktitle = {Adaptive Optics Systems VIII},
editor = {Laura Schreiber and Dirk Schmidt and Elise Vernet},
organization = {International Society for Optics and Photonics},
publisher = {SPIE},
pages = {1218514},
year = {2022},
doi = {10.1117/12.2628969},
URL = {https://doi.org/10.1117/12.2628969}
}

@Article{Wagner_2022,
  author    = {Wagner, Roland and Saxenhuber, Daniela and Ramlau, Ronny and Hubmer, Simon},
  journal   = {Astronomy and Computing},
  title     = {Direction dependent point spread function reconstruction for multi-conjugate adaptive optics on giant segmented mirror telescopes},
  year      = {2022},
  month     = {jul},
  pages     = {100590},
  volume    = {40},
  doi       = {10.1016/j.ascom.2022.100590},
  publisher = {Elsevier {BV}},
}

@article{Ono_2016,
author = {Yoshito H. Ono and Masayuki Akiyama and Shin Oya and Olivier Lardi\'{e}re and David R. Andersen and Carlos Correia and Kate Jackson and Colin Bradley},
journal = {J. Opt. Soc. Am. A},
number = {4},
pages = {726--740},
publisher = {Optica Publishing Group},
title = {Multi time-step wavefront reconstruction for tomographic adaptive-optics systems},
volume = {33},
month = {Apr},
year = {2016},
doi = {10.1364/JOSAA.33.000726},
}

@article{thiebaut2010fast,
  title={Fast minimum variance wavefront reconstruction for extremely large telescopes},
  author={Thi{\'e}baut, Eric and Tallon, Michel},
  journal={Journal of the Optical Society of America A},
  volume={27},
  number={5},
  pages={1046--1059},
  year={2010},
  publisher={Optical Society of America}
}

@inproceedings{tallon2010fractal,
  title={Fractal iterative method for fast atmospheric tomography on extremely large telescopes},
  author={Tallon, Michel and Tallon-Bosc, Isabelle and B{\'e}chet, Cl{\'e}mentine and Momey, Fabien and Fradin, Marie and Thi{\'e}baut, {\'E}ric},
  booktitle={Adaptive Optics Systems II},
  volume={7736},
  pages={354--363},
  year={2010},
  organization={SPIE}
}

@inproceedings{tallon2011performances,
  title={Performances of MCAO on the E-ELT using the fractal iterative method for fast atmospheric tomography},
  author={Tallon, Michel and B{\'e}chet, Cl{\'e}mentine and Tallon-Bosc, Isabelle and Le Louarn, Miska and Thi{\'e}baut, E and Clare, Richard and Marchetti, Enrico},
  booktitle={Adaptive Optics for ELTs II},
  year={2011}
}

@inproceedings{brunner2012optimal,
  title={Optimal projection of reconstructed layers onto deformable mirrors with fractal iterative method for AO tomography},
  author={Brunner, Elisabeth and B{\'e}chet, Clementine and Tallon, Michel},
  booktitle={Adaptive Optics Systems III},
  volume={8447},
  pages={1802--1810},
  year={2012},
  organization={SPIE}
}

@inproceedings{capasso2024morfeo,
  title={MORFEO at ELT: recent updates in the real-time computer design},
  author={Capasso, Giulio and Baruffolo, Andrea and Puglisi, Alfio T and Savarese, Salvatore and Lampitelli, Salvatore and Di Prospero, Chiara and Foppiani, Italo and Agapito, Guido and Busoni, Lorenzo and Dunn, Jennifer and others},
  booktitle={Software and Cyberinfrastructure for Astronomy VIII},
  volume={13101},
  pages={1328--1342},
  year={2024},
  organization={SPIE}
}

@article{StRa22,
author = {Bernadett Stadler and Ronny Ramlau},
title = {{Performance of an iterative wavelet reconstructor for the Multi-conjugate Adaptive Optics RelaY of the Extremely Large Telescope}},
volume = {8},
journal = {Journal of Astronomical Telescopes, Instruments, and Systems},
number = {2},
publisher = {SPIE},
pages = {1 -- 16},
year = {2022},
doi = {10.1117/1.JATIS.8.2.021503},
URL = {https://doi.org/10.1117/1.JATIS.8.2.021503}
}

@INPROCEEDINGS{StBiMaRa2021,
  author={Stadler, Bernadett and Biasi, Roberto and Manetti, Mauro and Ramlau, Ronny},
  booktitle={2021 21st International Conference on Computational Science and Its Applications (ICCSA)}, 
  title={Parallel implementation of an iterative solver for atmospheric tomography}, 
  year={2021},
  volume={},
  number={},
  pages={123-132},
  doi={10.1109/ICCSA54496.2021.00026}
}

@inproceedings{StBiMaRa19,
	author = {Stadler, Bernadett and Biasi, Roberto and Manetti, Mauro and Ramlau, Ronny},
	title = {{Feasibility of standard and novel solvers in atmospheric tomography for the ELT}},
	booktitle = {Proc. AO4ELT6},
	pages = {},
	year = {2019},
	doi = {}
}

@article{SaRa15, 
	author = {Saxenhuber, Daniela and Ramlau, Ronny}, 
	title  = {A Gradient-based method for atmospheric tomography},
	journal = {Inverse Problems and Imaging},
	year   = {2016},
	volume = {10},
	number = {3},
	pages = {781--805},
	doi = {http://dx.doi.org/10.3934/ipi.2016022}
}

@article{fusco2022key,
  title={Key wavefront sensors features for laser-assisted tomographic adaptive optics systems on the Extremely Large Telescope},
  author={Fusco, Thierry and Agapito, Guido and Neichel, Benoit and Oberti, Sylvain and Correia, Carlos and Haguenauer, Pierre and Plantet, C{\'e}dric and Pedreros, Felipe and Ke, Zibo and Costille, Anne and others},
  journal={Journal of Astronomical Telescopes, Instruments, and Systems},
  volume={8},
  number={2},
  pages={021514--021514},
  year={2022},
  publisher={Society of Photo-Optical Instrumentation Engineers}
}

@article{oberti2022super,
  title={Super-resolution wavefront reconstruction},
  author={Oberti, Sylvain and Correia, Carlos and Fusco, Thierry and Neichel, Benoit and Guiraud, Pierre},
  journal={Astronomy \& Astrophysics},
  volume={667},
  pages={A48},
  year={2022},
  publisher={EDP Sciences}
}

@inproceedings{cranney2024mavis,
  title={MAVIS: real-time wavefront estimation strategy},
  author={Cranney, Jesse and Agapito, Guido and Plantet, C{\'e}dric and Pinna, Enrico and Viotto, Valentina and Rigaut, Fran{\c{c}}ois and Doucet, Nicolas and Bernard, Julien and Gratadour, Damien and Taylor, Brian and others},
  booktitle={Adaptive Optics Systems IX},
  volume={13097},
  pages={1851--1861},
  year={2024},
  organization={SPIE}
}

@inproceedings{cranney2021optimising,
  title={Optimising wavefront sensing super-resolution in the control of tomographic adaptive optics},
  author={Cranney, Jesse and Guihot, Angus and De Dona, Jose and Rigaut, Francois},
  booktitle={2021 Australian \& New Zealand Control Conference (ANZCC)},
  pages={24--29},
  year={2021},
  organization={IEEE}
}

@inproceedings{MICADO24_psfr_offaxis,
author = {Matteo Simioni and Daniel Jodlbauer and Carmelo Arcidiacono and Andrea Grazian and Marco Gullieuszik and Elisa Portaluri and Benedetta Vulcani and Roland Wagner and Anita Zanella and Johanna Hartke and Tapio Helin and Hanindyo Kuncarayakti and Elena Masciadri and Fernando Pedichini and Roberto Piazzesi and Alessio Turchi and Piero Vaccari},
title = {{The MICADO first light imager for the ELT: off-axis performance of PSF reconstruction}},
volume = {13097},
booktitle = {Adaptive Optics Systems IX},
editor = {Kathryn J. Jackson and Dirk Schmidt and Elise Vernet},
organization = {International Society for Optics and Photonics},
publisher = {SPIE},
pages = {130977C},
year = {2024},
doi = {10.1117/12.3017375},
URL = {https://doi.org/10.1117/12.3017375}
}

@inproceedings{MICADO24_psfr_software,
author = {Andrea Grazian and Elisa Portaluri and Matteo Simioni and Carmelo Arcidiacono and Marco Gullieuszik and Johanna Hartke and Daniel Jodlbauer and Fernando Pedichini and Roberto Piazzesi and Piero Vaccari and Benedetta Vulcani and Roland Wagner and Anita Zanella},
title = {{The MICADO first light imager for the ELT: the PSF reconstruction software}},
volume = {13097},
booktitle = {Adaptive Optics Systems IX},
editor = {Kathryn J. Jackson and Dirk Schmidt and Elise Vernet},
organization = {International Society for Optics and Photonics},
publisher = {SPIE},
pages = {130977D},
year = {2024},
doi = {10.1117/12.3019757},
URL = {https://doi.org/10.1117/12.3019757}
}

\end{document}